\documentclass[a4paper,11pt]{article}
\usepackage{jcappub} 
\usepackage{lineno}
\usepackage{pifont}
\usepackage{array}
\usepackage{siunitx}
\usepackage{multirow}
\usepackage{graphicx}
\usepackage{subfig}
\usepackage{multicol}
\usepackage{comment}
\usepackage{dsfont}
\usepackage{soul}
\usepackage{txfonts}
\usepackage{color}

\arxivnumber{arXiv:2409.18714} 
\title{\boldmath Spectral Imaging with QUBIC: building astrophysical components from Time-Ordered-Data using Bolometric Interferometry}

\author[a]{M. Regnier,}
\author[a]{T.~Laclavere,}
\author[a]{J-Ch.~Hamilton,}
\author[b]{E.~Bunn,}
\author[a, c]{V.~Chabirand,}
\author[a]{P.~Chanial,}
\author[a]{L.~Goetz,}
\author[a]{L.~Kardum,}
\author[a]{P.~Masson,}
\author[d]{N.~Miron Granese,}
\author[e]{C.G.~Scóccola,}
\author[a, f]{S.A.~Torchinsky,}
\author[g, h]{E.~Battistelli,}
\author[i, j]{M.~Bersanelli,}
\author[g, h]{F.~Columbro,}
\author[g, h]{A.~Coppolecchia,}
\author[d]{B.~Costanza,}
\author[g, h]{P.~De Bernardis,}
\author[g, h]{G.~De Gasperis,}
\author[g, h]{S.~Ferazzoli,}
\author[k]{A.~Flood,}
\author[a]{K.~Ganga,}
\author[l, m]{M.~Gervasi,}
\author[a]{L.~Grandsire,}
\author[a]{A.~Huchet,}
\author[i, j]{E~.Manzan,}
\author[g, h]{S.~Masi,}
\author[i, j]{A.~Mennella,}
\author[n]{L.~Mousset,}
\author[k]{C.~O'Sullivan,}
\author[g, h]{A.~Paiella,}
\author[g, h]{F.~Piacentini,}
\author[a]{M.~Piat,}
\author[o]{L.~Piccirillo,}
\author[p]{E.~Rasztocky,}
\author[q, r]{M.~Stolpovskiy,}
\author[l, m]{M.~Zannoni}
\affiliation[a]{Université de Paris, CNRS, Astroparticule et Cosmologie, F-75006 Paris, France}
\affiliation[b]{University of Richmond, Richmond, USA}
\affiliation[c]{Ecole polytechnique, Institut Polytechnique de Paris, Palaiseau, France}
\affiliation[d]{Facultad de Ciencias Astronómicas y Geofísicas (Universidad Nacional de La Plata), Argentina}
\affiliation[e]{Cosmology and Theoretical Astrophysics group, Physics Department, FCFM, Universidad de Chile, Blanco Encalada 2008, Santiago, Chile}
\affiliation[f]{Observatoire de Paris, Université PSL, F-75013 Paris, France}
\affiliation[g]{Università di Roma - La Sapienza, Italy}
\affiliation[h]{INFN sezione di Roma, 00185 Roma, Italy}
\affiliation[i]{Universita degli studi di Milano, Italy}
\affiliation[j]{INFN sezione di Milano, 20133 Milano, Italy}
\affiliation[k]{National University of Ireland, Maynooth, Ireland}
\affiliation[l]{Università di Milano - Bicocca, Italy}
\affiliation[m]{INFN Milano-Bicocca, Italy}
\affiliation[n]{Laboratoire de Physique de l’École Normale Supérieure, ENS, Univ. PSL, CNRS, Sorbonne Univ., Univ. Par Cité, 75005 Paris, France}
\affiliation[o]{University of Manchester, UK}
\affiliation[p]{Instituto Argentino de Radioastronomía (CONICET, CIC), Argentina}
\affiliation[q]{International Space Science Institute (ISSI), Hallerstrasse 6, 8012 Bern, Bern, Switzerland}
\affiliation[r]{University of Bern, Hochschulstrasse 4, 3012 Bern, Switzerland}

\emailAdd{regnier@apc.in2p3.fr}

\abstract{   {The detection of \mbox{B-modes} in the CMB polarization pattern is a major issue in modern cosmology and must therefore be handled with analytical methods that produce reliable results. We describe a method that uses the frequency dependency of the QUBIC synthesized beam to perform component separation at the map-making stage, to obtain more precise results.}
   {We aim to demonstrate the feasibility of component separation during the map-making stage in time domain space. This new technique leads to a more accurate description of the data and reduces the biases in cosmological analysis.}
   {The method uses a library for highly parallel computation which facilitates the programming and permits the description of experiments as easily manipulated operators. These operators can be combined to obtain a joint analysis using several experiments leading to maximized precision. }
   {The results show that the method works well and permits end-to-end analysis for the CMB experiments, and in particular, for QUBIC. The method includes astrophysical foregrounds, and also systematic effects like gain variation in the detectors. We developed a software pipeline that produces uncertainties on tensor-to-scalar ratio at the level of $\sigma(r) \sim 0.023$ using only QUBIC simulated data.}}

\begin{document}
\maketitle
\flushbottom

\section{Introduction}

The cosmic microwave background (CMB) is one of the most important probes for understanding the primordial universe in modern cosmology. The COBE mission \cite{Fixsen_1996, Smoot_1999}  measured the black body nature and the intrinsic temperature anisotropies of the CMB establishing the Cold Dark Matter (CDM) model. Cosmology entered into the precision era with ground-based, balloon-borne, and satellite CMB experiments, such as Boomerang \citep{Masi_2002}, ACTPol~\citep{Thornton_2016}, WMAP \citep{Bennett_2013}, Planck \citep{planck2020}, and BICEP \citep{PhysRevLett.127.151301} that established the $\Lambda$CDM model, enhancing the precision of temperature anisotropies down to the cosmic-variance level throughout a wide multipole range.


Valuable information is not only encoded in the CMB temperature anisotropies but also in its polarization fluctuations. Current and planned CMB experiments aim at detecting primordial fluctuations by measuring the maps of the Stokes parameters I for intensity and Q~\&~U for polarization. The Q~\&~U maps 
do not allow a suitable description of the primordial fluctuations in terms of their scalar, vector, and tensor nature. It is, therefore, necessary to switch from Q/U to the well-known E and B modes, which describe the radial and rotational components of the polarization patterns respectively \cite{Zaldarriaga_2021}. In this context, primordial tensor fluctuations (primordial gravitational waves) produce \mbox{B-modes} at large angular scales ($\geq 0.5^{\circ}$), while scalar fluctuations do not produce \mbox{B-modes}. However, nonlinear effects that induce \mbox{B-modes} must also be considered. The most significant of these is weak lensing, which converts primordial \mbox{E-modes} into \mbox{B-modes} and peaks at intermediate and small scales. On the large angular scales, Galactic foreground contamination peaks at large angular scales.

The detection of primordial \mbox{B-modes}, present at large angular scales, would be indirect evidence for primordial gravitational waves and therefore a key observation in favor of the inflationary period. While a large number of experiments are currently working to measure the primordial \mbox{B-modes}, a positive detection remains elusive. 
The most stringent constraint on the tensor-to-scalar ratio is due to BICEP/Keck combining its data with WMAP and Planck, namely $r < 0.032$ with 95\% confidence~\citep{Tristram_2022} using cross-spectrum analysis through multi-components theory.

One of the most important issues in measuring the polarization signal of the CMB, and particularly the B-mode, for any experiment, is the ability to perform astrophysical component separation leaving as little separation residuals as possible. The observations of the CMB are naturally contaminated by foreground emissions for both temperature and polarization. Thermal dust and synchrotron emission are the main ones, respectively at high and low frequencies. Based on multi-frequency observation, component separation aims at cleaning the CMB maps from these foregrounds and is mandatory to extract unbiased CMB information within frequency maps describing the spectral energy distribution (SED) of the data. The faint signal related to the CMB polarization B-modes is entirely dominated by foregrounds. Therefore an accurate description of the foreground emission and its frequency dependence are required to constrain (or detect) the B-modes.

All ongoing experiments aiming at probing the primordial Universe through the CMB polarization B-modes put a significant emphasis on mitigating foreground contamination. This is done in an original manner with the Q~\&~U Bolometric Interferometer for Cosmology (QUBIC) which uses the technology of bolometric interferometry which combines the sensitivity of bolometers and the control of interferometric systematics (see ~\cite{2020.QUBIC.PAPER1, 2020.QUBIC.PAPER2,2020.QUBIC.PAPER3} while the instrument is described in \cite{2020.QUBIC.PAPER4, 2020.QUBIC.PAPER5, 2020.QUBIC.PAPER6, 2020.QUBIC.PAPER7, 2020.QUBIC.PAPER8} for a series of articles describing QUBIC instrumentation and scientific forecasts). Its interferometric nature gives it a special synthesized beam that is highly frequency-dependent, resulting in the ability to recover frequency-dependent 
information directly encoded in the full-bandwidth temporal data.

In this article, we maximally exploit this feature, presenting a method that combines map-making and component separation all at once from Time-Ordered-Data (TOD). This method allows reconstruction of the CMB polarization from the TODs directly without using frequency maps. The method has been constructed to produce the estimation of the astrophysical foreground using a mixing matrix as well as some systematic parameters such as the detector gains, similar to the method Commander3~\citep{refId0}. 

Section 2 is a summary of the spectral-imaging feature provided by Bolometric Interferometry. We present in Section 3 the simulations used for the reconstruction of the component maps. Sections 4 and 5 are devoted to defining the instrumental systematics parameters and the astrophysical components we reconstruct. Finally, section 6 presents the main results achieved with this innovative technique.

\section{Data acquisition and analysis with Bolometric Interferometry}
\label{spectral_imaging}

This work is a companion article to~\cite{fmm}, hereafter Frequency Map-Making (FMM), which describes the ability of Bolometric Interferometry (BI) to reconstruct sub-bands within the bandwidth of the instrument. Instead of reconstructing frequency maps, we aim here to directly perform component separation in the time domain. In this section, we will revisit the hypotheses and formalism described in FMM to apply them to joint map-making and astrophysical component separation.

\subsection{QUBIC synthesized beam}

The special characteristic of a Bolometric Interferometer such as QUBIC, the horn array, determines the shape of the synthesized beam. It is a superposition of interferometry fringe patterns, with fringe distance inversely proportional to frequency, exactly as in Young's double slit experiment. The obtained Point-Spread-Function (PSF), the sum of all fringes, is also frequency-dependent. The distance between the secondary and central peaks is inversely proportional to frequency. A more detailed description is given in ~\cite{2020.QUBIC.PAPER2}.

Due to the frequency dependence of the synthesized beam, the spectral information of the sky is spatially encoded on the focal plane, allowing the possibility to perform spectral imaging. This feature allows to retrieve spectral information inside the physical band of the instrument. In FMM, we demonstrated how to extract multiple frequency maps within one physical band. In this paper, we show that we can directly separate component maps at the TOD level by utilizing their spectral dependencies.

An important advantage of spectral imaging is that it is achieved at the data analysis stage, after data acquisition, enabling flexible sub-band division based on analytical needs, as demonstrated in~\cite{regnier2023identifying}.

\subsection{General principle}

As shown in FMM section~2.1, we introduce the operator $\vec{H}$ that describes how the instrument observes the sky. This acquisition matrix accounts for the well-known frequency variation of QUBIC's PSF (see section~2.2 in FMM) and several other instrumental acquisition features such as polarization modulation through the Half-Wave Plate or detector time constant. In this article, we no longer describe the sky as frequency observations but directly as a mixture of astrophysical components. This allows us to introduce the mixing matrix $\vec{A}$ that describes the power of each component at each frequency. The raw data can now be expressed using components instead of frequency observations through: 

\begin{equation}
    \vec{d} = \vec{H} \vec{A} \vec{c} + \vec{n}.
    \label{eq:data}
\end{equation}
The components vector $\vec{c}$ that contains components maps and the mixing matrix $\vec{A}$ , that describes their frequency evolution, will be properly defined in section \ref{sec: sky_model}.

The inverse-problem approach we adopt in FMM enables us to simulate the instrument data by observing a set of simulated components, denoted as $\vec{\Tilde{c}}$ (where the tilde indicates simulated data). Similar to the polychromatic model introduced in FMM (Eq.~11) for reconstructing sub-frequency maps, we use the following model to reconstruct sky component maps directly from the TOD:
\begin{equation}
   \vec{\tilde{d}} = \left( \sum_{k = 1}^{N_{\text{sub}}} \Delta \nu_{k} \mathcal{H}_{\nu_{k}} \mathcal{C}_{K_{k}} \vec{\Tilde{A}}_{\nu_{k}} \right) \cdot \vec{\Tilde{c}},
    \label{eq:spectral_im_model}
\end{equation}

In this model, the operators act on the component vector $\vec{\Tilde{c}}$ as follows:
\begin{enumerate}
    \item Simulate the mixing of components on the sky at frequency $\nu_{k}$ using $\vec{\Tilde{A}}_{\nu{k}}$.
    \item Convolve the resulting sky map at frequency $\nu_{k}$ with the instrument's angular resolution at that frequency using $\mathcal{C}_{K_k}$.
    \item Simulate the instrument's data acquisition process at frequency $\nu_{k}$ using $\mathcal{H}_{\nu{k}}$.
\end{enumerate}
The integral over frequencies is discretized into a sum, represented by $\Delta \nu_{k}$, which denotes the width of the discrete frequency intervals used in the model.

In appendix~\ref{sec: appendixB}, we explore the possibility of not doing those convolutions during reconstruction in order to speed up the calculations. The reconstruction will naturally fit the maps to an average resolution. We approximate analytically this resolution, allowing us to do the cosmological analysis, as described in section~\ref{sec:fit_r}.

\subsection{Adding external data}

In FMM, we emphasize the frequency dependence of QUBIC's PSF with respect to the wavelength. Still, we also report that the multiple peaks create biases on the edges of the reconstructed maps due to missing information related to some of the multiple peaks of the PSF pointing outside the mapped region. In an exactly similar way as in FMM, we regularize the reconstruction of these pixels using external data. For this purpose, we create a very general model including QUBIC and external data:

\begin{equation}
    \vec{H}_\text{Tot} = \begin{bmatrix}
        \vec{H}^{\text{QUBIC}} \\
        \vec{H}^{\text{Ext}} \\
        \end{bmatrix} = \renewcommand{\arraystretch}{1.5}\begin{bmatrix}
        \vec{H}^{\text{QUBIC}}_{150} \\
        \vec{H}^{\text{QUBIC}}_{220} \\
        \vec{H}^{\text{Ext}}_{\nu_1} \\
        \vdots \\
        \vec{H}^{\text{Ext}}_{\nu_N} \\
        \end{bmatrix},
        \label{eq:H_poly}
\end{equation}
where $\vec{H}^{\text{QUBIC}}$ is the application of Eq.~\ref{eq:spectral_im_model} for each of our focal planes (operating at 150 and 220 GHz respectively), $\vec{H}^{\text{Ext}}$ is a much simpler operator that only reads the external data sky maps. In the simulations presented in this article, our external dataset consists of simulated Planck data including noise following the published noise maps\footnote{{\tt http://pla.esac.esa.int/pla/\#maps}}. For real sky observations, the simulated external data will be replaced by observed sky maps.

As we have now a way to simulate time domain data from a guess set of sky components, we need to compare this simulation to the actual data minimizing a cost function (jointly for QUBIC and external data). It is very similar to FMM except that it explicitly depends upon foreground components $\vec{c}$ that are mixed at a given frequency through $\vec{A}$:
\begin{equation}
\begin{split}
    \chi^2_\text{Tot} &= \begin{bmatrix}
                    \vec{d}_\text{QUBIC} - \vec{\tilde{d}}_\text{QUBIC} \\
                    \vec{d}_\text{Ext} - \vec{\tilde{d}}_\text{Ext} \\
                \end{bmatrix}^T
                \begin{bmatrix}
                    \vec{N}^{-1}_\text{QUBIC} & 0 \\
                    0 &  \vec{N}^{-1}_\text{Ext} \\
                \end{bmatrix} 
                \begin{bmatrix}
                    \vec{d}_\text{QUBIC} - \vec{\tilde{d}}_\text{QUBIC} \\
                    \vec{d}_\text{Ext} - \vec{\tilde{d}}_\text{Ext} \\
                \end{bmatrix} \\
             &= \left( \vec{d}_\text{QUBIC} - \vec{H}^{\text{QUBIC}} \vec{\tilde{A}} \vec{\tilde{c}}\right)^T \vec{N}^{-1}_\text{QUBIC} \left( \vec{d}_\text{QUBIC} - \vec{H}^{\text{QUBIC}}\vec{\tilde{A}} \vec{\tilde{c}} \right) \\
             &~~~~~+
             \left( \vec{d}_\text{Ext} - \vec{H}^{\text{Ext}} \vec{\tilde{A}} \vec{\tilde{c}}\right)^T \vec{N}^{-1}_\text{Ext} \left( \vec{d}_\text{Ext} - \vec{H}^{\text{Ext}}\vec{\tilde{A}} \vec{\tilde{c}} \right) \\
             &= \chi^2_\text{QUBIC} + \chi^2_\text{Ext},
             \label{eq: sum chi2}
\end{split}
\end{equation}
where $\vec{N}^{-1}$ is the inverse noise covariance matrix weighting both experiments. In a similar manner as in FMM, we ensure that our method is nearly optimal by introducing weights for the external data. We also set the external data weights to zero within the QUBIC patch in order to minimize systematics from external data in the QUBIC patch.

\section{Sky model}\label{sec: sky_model}

Using this framework, we will reconstruct component maps through the above instrumental model, directly in the time domain. We consider in that case several sky components as the Cosmic Microwave Background fluctuations which is the most interesting signal for QUBIC. Astrophysical foregrounds are contaminating data, especially the thermal dust emission at high frequency and synchrotron emission at low frequency. Monochromatic polluting emissions can be found in QUBIC data such as the Carbon Monoxide (CO) line. Note that all the components are simulated with PySM\footnote{{\tt https://pysm3.readthedocs.io/en/latest}} software.

\subsection{Cosmic Microwave Background (CMB)}

We simulate the CMB fluctuations according to the Planck best-fitted cosmological parameters outlined as presented in the publication by \cite{planck2020}. The well-established flat spectrum of the CMB, as demonstrated by \cite{Fixsen_1996}, allows us to set the normalization of the CMB contribution within the mixing matrix to a value of one.

The cosmological parameters related to B-modes are fixed at $r = 0$ (tensor-to-scalar ratio) and $A_{lens} = 1$ (gravitational lensing residual) unless explicitly mentioned. A more detailed article on forecasts on the tensor-to-scalar ratio with QUBIC using the technique presented here and in FMM is anticipated for the near future.

\subsection{Thermal dust}
\label{sec:thermaldust}

Within the interstellar medium (ISM), tiny grains are present and receive heat from nearby stars. These minuscule grains cluster together to create clouds with temperatures approximately around $T_d = 20^{\circ}K$, emitting partially polarized radiation mainly in the sub-millimeter (sub-mm) wavelength range. This emitted radiation significantly interferes with high-frequency observations of the Cosmic Microwave Background (CMB), particularly when seeking to detect B-modes. A commonly used approach to represent the SED of this contaminant is through the modified black body (MBB) model:
\begin{equation}
    f_d^{\,\beta_d}(\nu) = \frac{e^{\frac{h\nu_0}{kT_d}}-1}{e^{\frac{h\nu}{kT_d}}-1}\left(\frac{\nu}{\nu_0}\right)^{1+\beta_d} \cdot \frac{f_\text{CMB}(\nu_0)}{f_\text{CMB}(\nu)},
    \label{eq:dust_model}
\end{equation}
with $f_\text{CMB}$ a conversion factor to express the maps in units $\si{\mu\K}_\text{CMB}$:
\begin{equation}
    f_\text{CMB}(\nu) = \frac{e^{\frac{h\nu}{k T_\text{CMB}}} \left(\frac{h\nu}{k T_\text{CMB}}\right)^2}{\left(e^{\frac{h\nu}{k T_\text{CMB}}}-1\right)^2}.
\end{equation}

The model, as encapsulated by Eq.~\ref{eq:dust_model}, shows two essential free parameters. These parameters, namely the temperature of the constituent dust grains (referred to as $T_d$) and the slope characterizing the spectral distribution, known as the spectral index $\beta_d$, play a main role in characterizing the dust component. It's important to note that for this study, the temperature of the dust is held constant at $T_d = 20^{\circ}K$ while $\nu_0$ is a reference frequency.

Furthermore, variations of the spectral indices across the sky are simulated using the PySM software, as shown in the publication by \citep{Thorne_2017}. The procedure for fitting these spectral indices is described in section~\ref{sec:varying_spectral_indices}. 
Note that one expects the dust spectral index to vary across the sky. To simulate realistic behavior, we use two distinct levels of pixelization: one for the component maps, set at $N_{\text{side}} = 256$, and a coarser one for the spectral indices. This coarser pixelization speeds up calculations and provides a higher signal-to-noise ratio while still adhering to a MBB distribution with a varying spectral index across the sky.

\subsection{Synchrotron emission}

The partially polarized synchrotron emission, originating from free electrons interacting with the Galactic magnetic field, exerts minimal influence on QUBIC, as illustrated in Fig.~12 in ~\cite{2020.QUBIC.PAPER1}. Consequently, the determination of the synchrotron spectral index primarily relies on external data sources, as discussed in section~\ref{data}. This study is primarily dedicated to performing component separation during the map-making phase for QUBIC rather than reanalyzing existing datasets.
\begin{equation}
    f_s^{\,\beta_s}(\nu) = \left( \frac{\nu}{\nu_0} \right)^{\beta_s} \cdot \frac{f_\text{CMB}(\nu_0)}{f_\text{CMB}(\nu)}.
    \label{eq:sync_model}
\end{equation}

We represent this emission with a power law, defined by Eq.~\ref{eq:sync_model}. This emission possesses its unique spectral index, denoted as $\beta_s$.\\

The components vector $\vec{c} = \left(\vec{s}_\text{CMB},\, \vec{s}_\text{dust},\, \vec{s}_\text{sync}\right)$ is the combination of the CMB, the dust, and the synchrotron. The mixing matrix $\vec{A}_\nu$ acts on $\vec{c}$ to produce the mixed sky at frequency $\nu$:
\begin{equation}
    \vec{A}_\nu \vec{c} = \vec{s}_\text{CMB} + f_d^{\,\beta_d}(\nu) \cdot \vec{s}_\text{dust} + f_s^{\,\beta_s}(\nu) \cdot \vec{s}_\text{sync}.
\end{equation}

\subsection{CO line}
\label{co_line}

In the context of this study, we present a promising avenue for the QUBIC instrument, involving the reconstruction of Carbon Monoxide (CO) line emissions. One of this particular spectral line was observed in the 217 GHz band of Planck~\citep{co_line_planck} at the particular frequency $\nu_{\text{CO}} = 230.538$ GHz and should therefore be seen in the 220 GHz band of QUBIC that spans 192.5 to 247.5 GHz. To facilitate our analysis, we incorporate a monochromatic representation of this emission within our model, a choice justified by the frequency-dependent characteristics of QUBIC's synthesized beam. This emission is probably not very polarized, as it hasn't been detected yet in polarization from current data~\citep{co_line_planck}. Our method nevertheless allows us to reconstruct this emission in terms of both intensity and polarization, should it be so. We therefore assume a 1\% polarization level for the CO line in our simulations for the sake of demonstrating our ability to reconstruct such a line in polarization with our technique.

Our approach enables the extraction of specific frequency components from raw data, a capability arising from the frequency-dependent nature of QUBIC's synthesized beam. To appreciate the significance of this emission and its associated challenges, we reference the empirical results obtained by the Planck satellite mission, as detailed in~\cite{co_line_planck}. In particular, an insightful perspective can be gained by examining Fig.~4 within this reference, which illustrates the relatively faint nature of the CO line emission. We describe CO as a monochromatic emission in our model by adding to the former operator $H_\text{Tot}$ (Eq.~\ref{eq:H_poly}), an additional acquisition operator $H_{\text{CO}}$:
\begin{equation}
    \vec{H}_{\text{Tot}} \rightarrow \vec{H}_{\text{Tot}} + \vec{H}_{\text{CO}},
    \label{eq:hco}
\end{equation}
where $H_{\text{CO}}$ only operates at the emission line frequency $\nu_{CO}=230.538 \text{ GHz}$ in that case and therefore uses a monochromatic version of the synthesized beam.
 
While QUBIC is not primarily designed for the reconstruction of such emissions, the spectral sensitivity from Bolometric Interferometry along with our inverse problem approach makes it straightforward to implement reducing the impact of the dilution within the large bandpass that an imager would experience.

\section{Data overview}
\label{data}

Our approach combines different sets of information to figure out the astrophysical elements. These sets of information can be quite complex, like the ones from QUBIC, or simpler, from previous experiments like Planck. This part explains the data we use in our approach.

\subsection{QUBIC: 150 \& 220 GHz}

In our study, QUBIC observations are crafted using our component-based acquisition model. These simulations are carried out over a duration equivalent to three years of cumulative observation time. The simulations are generated to encapsulate a composite sky, combining realizations of the Cosmic Microwave Background (CMB) and various astrophysical foregrounds.

\begin{table}[h!]
    \renewcommand{\arraystretch}{1.2}
    \begin{center}
        \caption{\label{tab: main_parameters}Main parameters of QUBIC Full Instrument (FI).}
        \begin{tabular}{p{6cm} m{2cm} m{2cm}}
            \hline
            \hline
            {\bf Parameters}\\
            \hline
            Frequency channels [GHz] \dotfill & 150 & 220 \\
            Bandwidth [GHz] \dotfill & 37.5 & 55 \\
            Effective FWHM [$^{\circ}$] (arcmin) \dotfill & 0.39 (23.4) & 0.27 (16.2) \\
            Number of detector & $992$ & $992$ \\
        \end{tabular}
    \end{center}
\end{table}

It is essential to note that the Full QUBIC instrument (see Table~\ref{tab: main_parameters}), as envisioned, will comprise two focal planes, each equipped with 992 detectors. Each focal plane is set to work at 150 GHz and 220 GHz respectively. These detectors yield two sets of Time-Ordered Data (TODs), and these TODs will play a crucial role in our final data analysis. As shown in the FMM paper, the ability of QUBIC to perform spectral-imaging allows us to increase the spectral resolution to have better mitigation of astrophysical foregrounds. 

\subsection{Planck Low Frequency Instrument (LFI)}

In our research, we relied on the most reliable constraints currently available from the Planck satellite. The low-frequency instrument (LFI) on Planck made observations across the entire sky at three distinct frequencies, where polarized synchrotron emission dominates. The process of fitting the spectral index of synchrotron emissions primarily hinges on external data sources. We do not anticipate achieving superior results in this aspect compared to the Planck LFI's performance on low-frequency observations.

\subsection{Planck High Frequency Instrument (HFI)}

In the same way, we simulated the high frequency instrument of Planck (HFI) using the parameters in the Table~\ref{tab_hfi_params}. That part of the external data is more important in our case because QUBIC will directly observe this frequency range.
\begin{table}[h!]
    \renewcommand{\arraystretch}{1.2}
    \begin{center}
        \caption{\label{tab_hfi_params}Planck HFI main parameters.}
        \begin{tabular}{p{6cm} m{1cm} m{1cm} m{1cm} m{1cm}}
            \hline
            \hline
            {\bf Parameters}\\
            \hline
            Frequency channels [GHz] \dotfill & 100 & 143 & 217 & 353\\
            Bandwidth [GHz] \dotfill & 33 & 47.1 & 71.6 & 116.5\\
            Effective FWHM [arcmin] \dotfill & 9.66 & 7.22 & 4.90 & 4.92 \\
            Sensitivity depth [$\mu\text{K.arcmin}$] \dotfill & 118 & 70.2 & 105. & 439\\
        \end{tabular}
    \end{center}
\end{table}

The most important HFI frequency range for this study is the $353 \text{ GHz}$ band, which is very important to constrain the thermal dust as it provides information on the dust SED over a larger frequency range. Conversely, the sampling of the SED done by QUBIC allows us to access higher frequency resolution ($\Delta \nu / \nu \approx 0.05$), driven by the geometry of the apertures of the interferometer~\citep{2020.QUBIC.PAPER2}. The $217$ GHz band of Planck also contains CO line emission described in section \ref{co_line}, which is highly diluted inside the thermal dust signal. This particular line is also in the QUBIC 220 GHz data. In this way, by including in the model a monochromatic description of that emission, we should be able to reconstruct it. To study that possibility, we chose to move the sky coverage of QUBIC to the galactic plane to observe a sizeable signal (see section~\ref{sec: monochromatic line recon}).

\section{Components map-making algorithm}

In FMM, we provide an introduction to the fundamentals of the inverse map-making problem and its solution. However, our primary focus here is to present the algorithm employed for the direct separation of astrophysical components. This approach eliminates the necessity for frequency map projection. With this method, we harness the sensitivity of a broad-spectrum instrument and gain advanced control over astrophysical foregrounds across the entire frequency range.

\subsection{Minimization process}
\label{sec:review}

\begin{figure}
    \centering
    \includegraphics[scale=0.6]{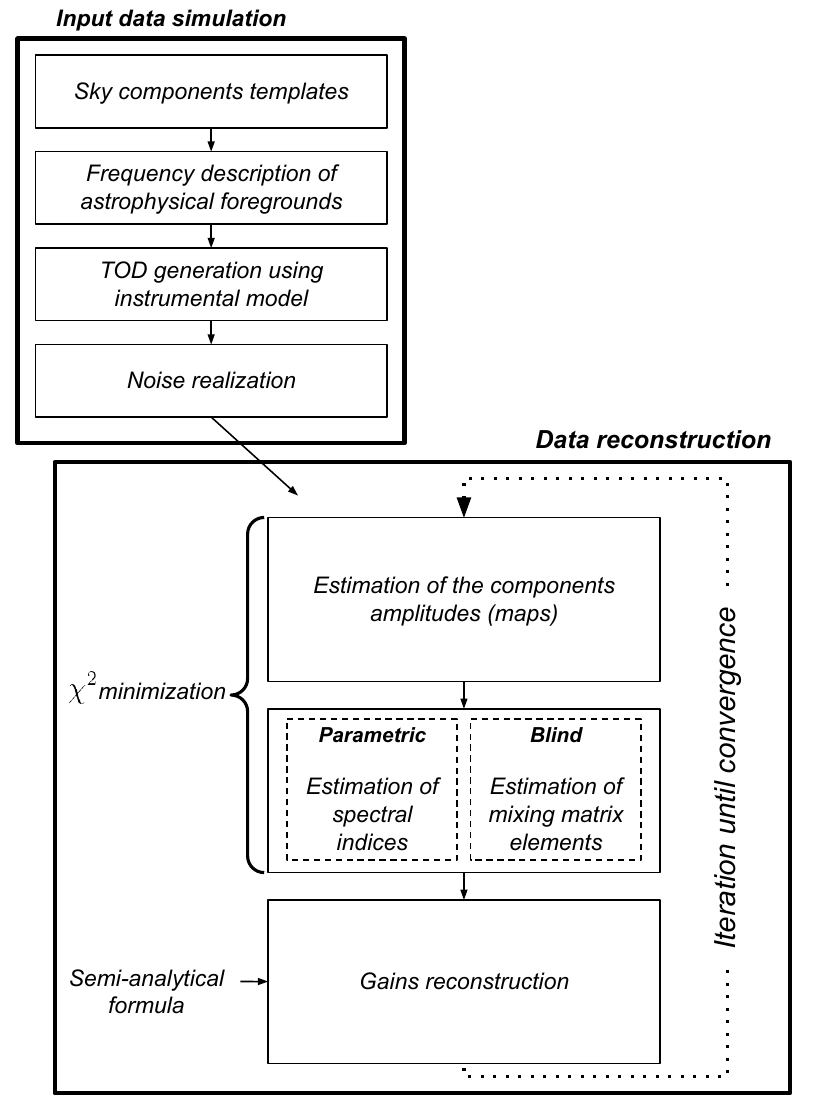}
    \caption{Illustration of the Components Map-Making.}
    \label{fig: illustration_cmm}
\end{figure}

The algorithm employed in this study performs component separation in the time domain by solving linear map-making equations iteratively. The presence of astrophysical foregrounds needs knowledge of their spectrum associated with spectral indices introduced in section~\ref{sec: sky_model}. Those parameters are non-linear, which makes the minimization process more complex. We decided to perform alternate minimization by updating free parameters. We will describe in this section each stage of this alternate minimization process. An illustration of the algorithm is shown in Fig.~\ref{fig: illustration_cmm}.

\subsubsection{Preconditioned Conjugate Gradient (PCG)}
\label{sec:pcg_maps}
In this paper and in the FMM, we use the PCG to solve our linear system $\nabla\chi^2 (\vec{\Tilde{c}}) = 0$, equivalent to $(\vec{H}\vec{\Tilde{A}})^T \vec{N}^{-1} \vec{H}\vec{\Tilde{A}} \vec{\Tilde{c}} = (\vec{H}\vec{\Tilde{A}})^T \vec{N}^{-1} \vec{d}$. Also described in FMM, we use this method to solve a system with many unknowns as amplitudes of the components. This method allows us to quickly converge to the optimal solution that minimizes our residuals by computing an optimal direction.

The package\footnote{PyOperators used massively parallel libraries, developed by P. Chanial ({\tt https://github.com/pchanial/pyoperators})} used for this method controls allocated memory and massive parallelization to the benefit of computing resources. Detectors are parallelized, allowing them to be processed independently. Each step of the algorithm is parallelized, which means resources are better distributed. 

This approach determines an efficient path for minimizing the cost function, which is established based on our acquisition model as described in section~\ref{spectral_imaging}. Our model, inclusive of the mixing matrix $\vec{A}$, presupposes consistent spectral indices and systematic parameters for a specific component vector $\vec{c}$. This framework enables us to perform the PCG method iteratively, alternating between fitting spectral indices and instrumental systematics multiple times.



Eq.~\ref{eq: sum chi2} shows that the total $\chi^2_\text{tot}$ is a sum of contributions coming from different experiments, QUBIC and Planck here. Note that we don't use Planck TOD but published sky pixels, weighted by their noise variance in the noise covariance matrix. We can control the weight of each experiment in order to fully benefit from external data. We force the acquisition to assign zero weight to external data inside the QUBIC patch in order to avoid possible contamination of external data in terms of instrumental systematic. This method produces a joint analysis between many experiments by their description of taking data.

\subsubsection{Gain estimation: Semi-analytical solution}
\label{sec: gain_fit}

We introduce in our method the possibility of estimating systematics directly during the minimization. As an example, we model the gain for each detector. In the future, we will add more refined effects. In the simulations, we consider gains distributed according to a normal distribution centered on 1 with a given standard deviation. QUBIC has 1 focal plane with 248 detectors for the Technical Demonstrator (TD) and will have 2 focal planes with 992 detectors each for the Full Instrument (FI). Perfectly intercalibrated data are created by Eq.~\ref{eq:data}. In practice, we measure: 
\begin{equation}
    \vec{d} = \vec{G} \vec{D} + \vec{n},
    \label{eq:data_g}
\end{equation}
where $\vec{D} = \vec{H} \vec{A} \vec{c}$ and $\vec{G}$ are the unknown intercalibrations. Each detector has its independent intercalibration factor, which we normalize with respect to a reference detector, which by definition has a gain of 1. The whole computation is in appendix~\ref{appendixA} where the final result is given by:
\begin{equation}
    \vec{G}_i = \frac{\vec{D}_i \vec{N}^{-1} \vec{d}}{\vec{D}_i \vec{N}^{-1} \vec{D}_i},
    \label{eq:gain}
\end{equation}
for the $i$-th iteration. Intercalibration factors are computed at each iteration with a new set of components and spectral indices. This method enables us to calculate these systematics quickly and easily using our simulated data. Further adjustments are required to better assess and incorporate a wider range of systematic influences into the model.

\begin{figure}
    \centering
    \includegraphics[scale=0.55]{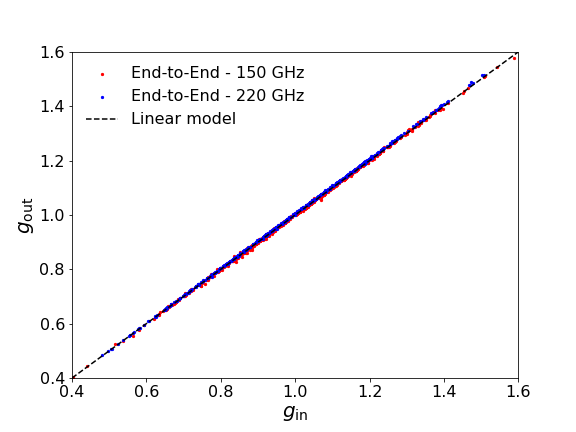}
    \caption{Reconstructed gain at 150\,GHz (red) and at 220\,GHz (blue) for a single realization of CMB + dust (model \textbf{d0}) sky model.}
    \label{fig: gain}
\end{figure}

We performed a single realization to illustrate our purpose and show the reconstruction of the gain of each detector in Fig.~\ref{fig: gain}. The input vs output plot shows a linear relation between the two, demonstrating a reliable reconstruction of the detector gain.

\subsubsection{Spectral indices fit}

We will now focus on characterizing the spectral behavior of astrophysical foregrounds. As said before, they are fit separately from gains and foregronuds maps, through the nested minimization presented in Fig.~\ref{fig: illustration_cmm}. The emission of those foregrounds is completely described through models introduced in section~\ref{sec: sky_model}, and may or may not consider the spatial variation of their values across the sky. In the same way as for the gains (see section~\ref{sec: gain_fit}), we need to write down a cost function dedicated to these parameters on a given line-of-sight (LOS):
\begin{equation}
    \chi^2_i(\vec{\beta}) = \left( \vec{d} - \vec{G}_i \vec{H} \vec{\Tilde{A}}_i(\vec{\beta}) \vec{\Tilde{c}}_i \right)^T \vec{N}^{-1} \left( \vec{d} - \vec{G}_i \vec{H} \vec{\Tilde{A}}_i(\vec{\beta}) \vec{\Tilde{c}}_i \right).
    \label{eq:chi2_indice}
\end{equation}
We set up a cost function based on section~\ref{spectral_imaging} but parametrized by spectral indices assuming that the components and gains $\vec{c}_i$ and $\vec{G}_i$ respectively for the $i$-th iteration are fixed due to the iterative scheme of the method (see the data reconstruction part in Fig.~\ref{fig: illustration_cmm}). This cost function is then minimized to find the maximum probability for a parameter using the Python {\tt scipy.optimize} package\footnote{{\tt https://github.com/scipy/scipy/tree/main/scipy\\ /optimize}} instead of PCG which is not efficient for non-linear parameters such as the spectral indices. 

Two parametric setups are available for characterizing the foreground emissions.  One entails employing a single, constant spectral index across the entire sky, while the other involves varying these parameters according to the pointing direction. The latter option is more faithful to real-world data but is also more difficult to implement. Our approach accommodates both configurations and can be sped up by exploiting parallelization techniques. As explained before, we use coarser pixelization for spectral indices than for the actual sky components maps.

\subsubsection{Reconstruction of the tensor-to-scalar ratio $r$}
\label{sec:fit_r}

The primary objective of this investigation is to establish a standardized approach for conducting component separation during the map-making process, aiming to estimate the cleanest components. In order to derive cosmological and astrophysical parameters, we employ the NaMaster\footnote{{\tt https://namaster.readthedocs.io/en/latest/}} code~\citep{Alonso_2019}, conducting auto-spectra for each component and cross-spectra for combinations of components. 

We perform a Monte-Carlo simulation where the CMB realizations are all similar (sample variance is treated analytically at the power spectrum level) and only noise vary from one realization to another. We average out the reconstructed components maps among these realization to achieve maps where the noise is reduced to a negligible contribution. Our average components maps therefore contain the ``true'' input component signal convolved to our resolution as well as possible foreground separation residuals. The power spectra of these maps writes:
\begin{align}
    \textbf{D}^{BB}_{\ell,\, \text{exp}} &= \textbf{D}^{BB}_{\ell,\, \text{model}} + \textbf{D}^{\text{sys}}_{\ell}
\end{align}
where $\textbf{D}^{BB}_{\ell,\, \text{model}}$ is the theoretical power spectrum and $\textbf{D}^{\text{sys}}_{\ell}$ is the systematic bias due to potential foreground residuals.

We write the likelihood on $r$ and foreground parameters using a Gaussian approximation~\citep{Hamimeche_2008}:
\begin{equation}
    -2 \ln \mathcal{L}(r, A_d, \alpha_d) = \left( \textbf{D}^{BB}_{\ell,\, \text{exp}} - \textbf{D}^{BB}_{\ell,\, \text{model}} \right)^T \textbf{N}_{\ell,\, \ell}^{-1} \left( \textbf{D}^{BB}_{\ell,\, \text{exp}} - \textbf{D}^{BB}_{\ell,\, \text{model}} \right),
    \label{chi2}
\end{equation}

where $\textbf{D}^{BB}_{\ell,\, \text{exp}}$ and $\textbf{D}^{BB}_{\ell,\, \text{model}}$ represent the measured and theoretical cross-spectra and auto-spectra of CMB and astrophysical foregrounds. The covariance matrix $\textbf{N}_{\ell,\, \ell}$ is the sum of two matrices. The first is the noise covariance matrix, obtained from the residual maps in our Monte-Carlo simulation where foreground separation biases are negligible ({\bf d0} model). In the second, the sample variance is added analytically following the derivation in FMM. The inverse covariance matrix is then corrected for the statistical bias arising from the finite number of realizations in our Monte-Carlo following~\citep{Taylor_2013}. The tensor-to-scalar ratio $r$ is simultaneously fitted with foreground parameters to leverage the anti-correlation between noise in different components arising from component separation. The error $\sigma(r)$ is computed by taking the standard deviation of the obtained Monte-Carlo Markov Chains.

\begin{table}
    \renewcommand{\arraystretch}{1.5}
    \setlength\tabcolsep{4pt}
    \begin{center}
        \caption{Parameters prior for the pipeline.\label{tab:priors}}
        \begin{tabular}{cccccc}
            \hline
            \hline
            {\bf Parameters} & $r$ & $A_d$ & $\alpha_d$ & $A_s$ & $\alpha_s$ \\
            \hline
            Prior \dotfill & $[-1, 1]$ & $[0, +\infty]$ & $[-\infty, 0]$ & $[0, +\infty]$ & $[-\infty, 0]$
        \end{tabular}
    \end{center}
\end{table}

Table~\ref{tab:priors} outlines the priors employed to constrain the post-component separation parameters. Despite the lack of physical significance for a negative value of the tensor-to-scalar ratio, incorporating such a range is deemed essential to maintain sensitivity to a potential source of bias and the robustness of our methodology. The parameter $A_\text{lens}$ is fixed at~1 during the fitting process as well as the reionization optical depth  $\tau = 0.054$~\citep{planck2020}.

\section{Results}

In the previous sections, we described our astrophysical component reconstruction algorithm based on simultaneous map-making and component separation. This section is dedicated to applying the algorithm to several simulated datasets, each corresponding to different assumptions regarding the complexity of the foregrounds. 
Section~\ref{parametric} explores the performance of the parametric version of our component map-making algorithm using simulations where the foregrounds are well described by an underlying SED. We first address the simple case of constant spectral indices over the sky in section~\ref{sec:constant_spectral_index} and then explore in section~\ref{sec:varying_spectral_indices}  the case of spatial variations of the spectral index across the sky. In section~\ref{sec:blind}, we explore the more complex situation where there is no global underlying SED model that describes the sky accurately, requiring a blind, non-parametric, version of our component map-making algorithm.

\subsection{Parametric component separation map-making}\label{parametric}
\subsubsection{Constant spectral index}
\label{sec:constant_spectral_index}

\begin{figure*}
    \centering
    \resizebox{\hsize}{!}{\includegraphics[scale = 0.9]{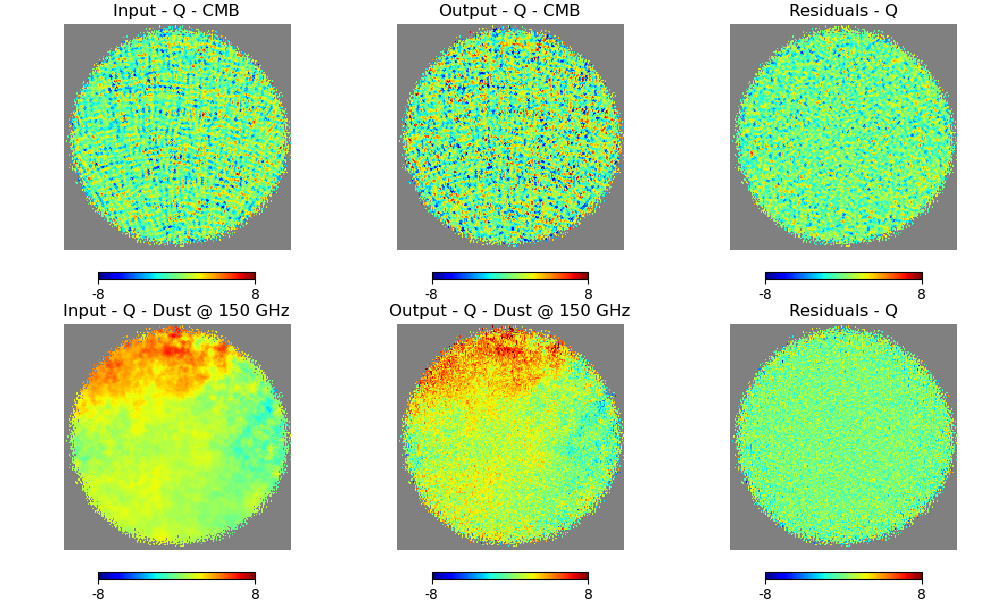}}
    \caption{Reconstruction of the Q Stokes parameter, in $\si{\mu\K}_\text{CMB}$, for both components on the QUBIC patch (centered on $15^{\circ}$ radius sky patch at $\text{RA}=0^{\circ}$, $\text{DEC}=-57^{\circ}$) assuming constant spectral index across the sky. The assumed data are QUBIC $150$ GHz + $220$ GHz + Planck HFI. Each row represents a component, and columns show the input, output, and residuals from left to right.}
    \label{fig:components}
\end{figure*}

As a first simple example, we can study component reconstruction assuming primordial fluctuations contaminated with thermal dust characterized by an MBB law with constant spectral index across the sky, known as the \textbf{d0} model\footnote{{\tt https://pysm3.readthedocs.io/en/latest/models.html}}. As presented in section~\ref{sec:thermaldust}, it can be described by two spectral parameters:~~The dust spectral index $\beta_d$ and the black body temperature \mbox{$T_d = 20$~K}. We include synchrotron emission in the raw data, but due to QUBIC’s frequency range, we do not expect to reconstruct this emission accurately enough to include it in our free components. 

\begin{figure}
    \centering
    \includegraphics[scale=0.5]{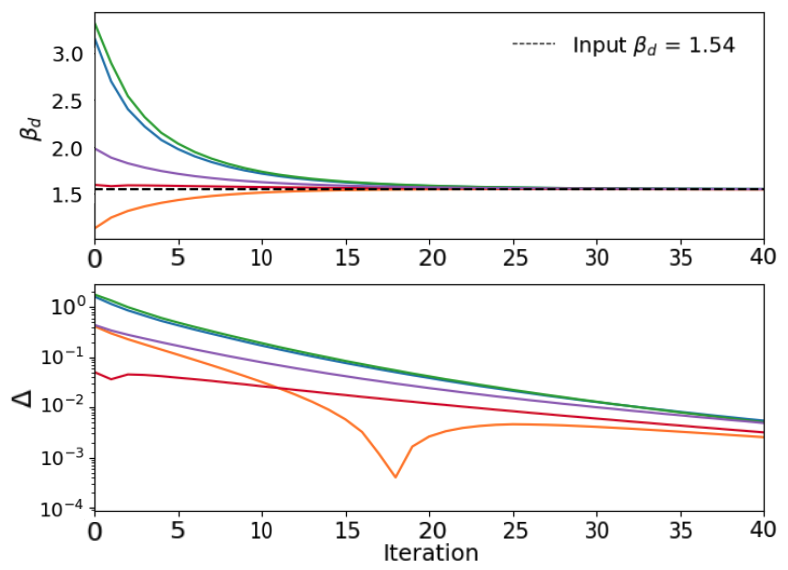}
    \caption{Convergence of the spectral index as a function of the number of iterations. Each color shows different noise realizations using different starting points. Note that despite the widely spread initial values, the algorithm converges. The bottom plot shows the residual with respect to the input value of the simulation defined by $\Delta = |\beta_{\text{input}} - \beta_{\text{output}}|$.}
    \label{fig:convergence_index}
\end{figure}

We generate simulated Time-Ordered Data using eight acquisitions to integrate the sky across the physical bandwidth of the instrument, taking into account the frequency evolution of the synthesized beam in our bolometric interferometer. As detailed in Section~\ref{sec:review}, we perform the $\chi^2$ minimization through an alternate estimation involving sky amplitudes and spectral index. The converged maps are displayed in Fig.~\ref{fig:components} showing the input, output, and residuals from left to right, respectively. Convergence is achieved when the parameters show minimal variation with further iterations. We have checked the convergence of the reconstructed components by computing the standard deviation of the maps within the QUBIC patch. We have measured that convergence is reached for a few hundred of iterations, for which the maximum number has been set at a thousand. The convergence for the spectral index is shown in Fig.~\ref{fig:convergence_index}, where absolute residuals $\Delta = |\beta_{\text{input}} - \beta_{\text{output}}|$ are presented. Simulations with different starting points for spectral indices yield an unchanged converged value, indicating that the results of our method are independent of the starting point. However, a suitable initialization accelerates the convergence toward the optimal solution.

The method used in this paper benefits from all the classical ways of parallelizing operations and spreading the work over several processors. We further improve convergence speed for the component maps by using Planck component maps as an initial guess~\citep{2020} except for the CMB component which is initialized to zero for Q/U Stokes parameters. The computations done in this article take about a day in calculation speed for each dataset, using 4 cores with 4 CPUs each. In each core, the focal planes are divided using MPI~\citep{DALCIN20051108} into several subsets of detectors that are jointly analyzed.

Therefore, we expect that synchrotron residuals in clean CMB maps will induce a bias on $r$, although at a small level. The CO line is not included in this simple case as it is mainly visible near the Galactic plane. Results on the CO line reconstruction are shown in section~\ref{sec: monochromatic line recon}.

\begin{figure}
    \centering
    \includegraphics[scale=0.4]{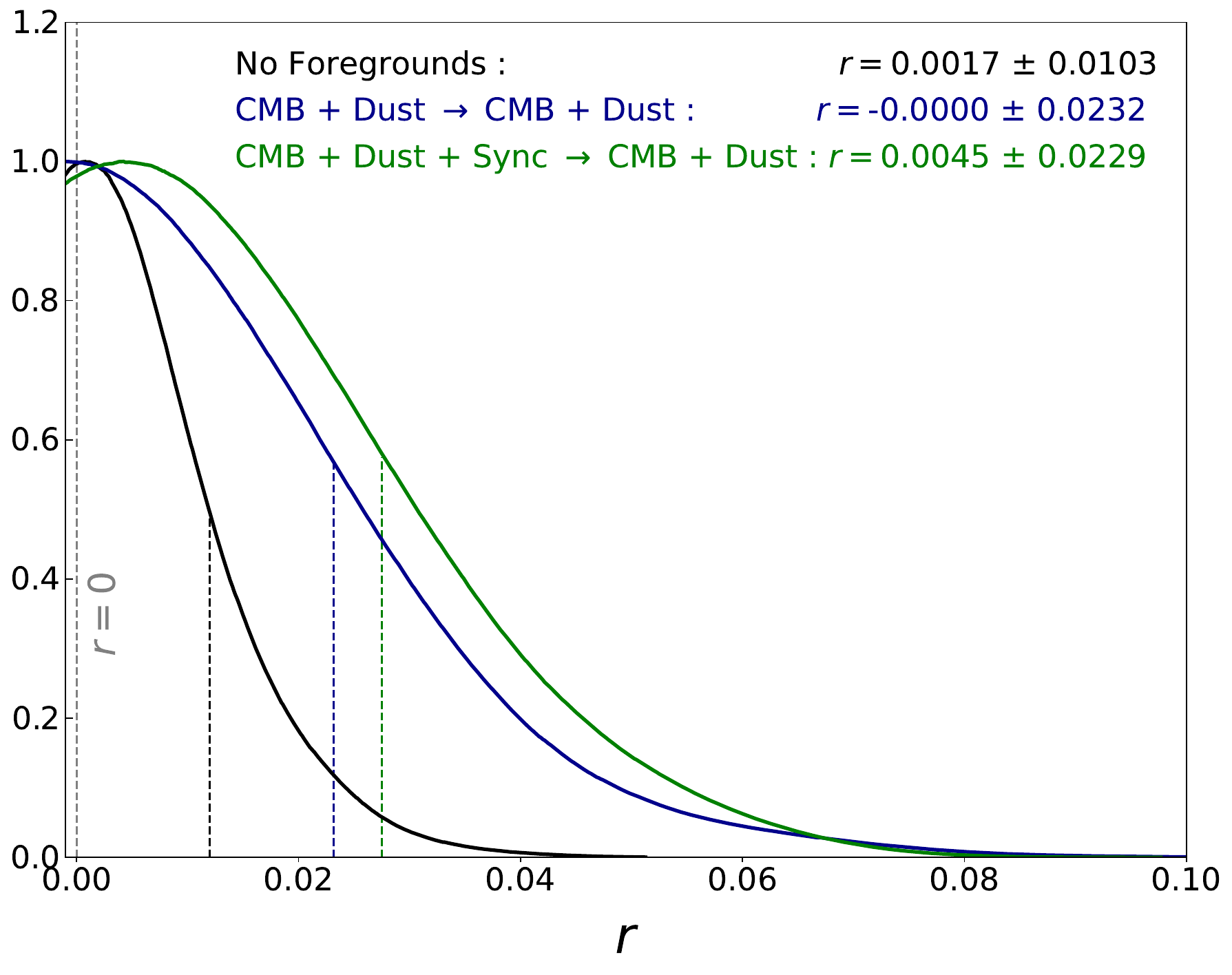}
    \caption{Posterior likelihood on tensor-to-scalar ratio $r$ assuming \textbf{d0} model.}
    \label{fig:cosmo_r_constant}
\end{figure}

We conducted a Monte-Carlo analysis to reconstruct the tensor-to-scalar ratio $r$ along with its associated error by accounting for the cosmic variance described by Eq.~3.2 of \cite{2020.QUBIC.PAPER1}. Simultaneously, we applied the same methodology to deduce astrophysical parameters, as elaborated in section~\ref{sec:fit_r}. The results of these analyses are visually depicted in Fig.~\ref{fig:cosmo_r_constant}, showcasing the posterior likelihood distribution for $r$. We explored two scenarios: The first, represented by the blue curve, where the synchrotron was not included in the input data; and the second, in green, incorporating the synchrotron. The unaccounted-for synchrotron introduces a bias in the reconstruction of $r$ around $r \approx 0.0045$ which cannot be reduced with QUBIC alone due to its frequency range. Even though this level of bias is not significant compared to QUBIC's precision on $r$, we checked that this bias was persistent with the number of realizations. Our simulations indicate that QUBIC's precision in measuring B-mode is anticipated to be $\sigma(r) = 0.0229$ after component separation. This process introduces a degradation factor $2.3$ when compared to a scenario involving only CMB, where $\sigma(r) = 0.0103$.

\subsubsection{Varying spectral indices}
\label{sec:varying_spectral_indices}

We now investigate a more realistic model where the value of the spectral index depends on the line-of-sight (LOS), based on \cite{planck_beta} and known as the \textbf{d1} model.

Rather than using a single spectral index for the entire sky, we characterize the emission of each component along the LOS. This approach results in a sky map of spectral indices, denoted as $\beta$, for each astrophysical foreground. To mitigate computational constraints, it is best to opt for a reduced pixelization of the spectral indices map $\vec{\beta}$ compared to that of the component map, assuming independence of spectral index pixels from one another. However, during the reconstruction phase, the mismatch between the chosen pixelization and the actual spatial variations of the spectral indices introduces foreground residuals in the clean CMB maps, consequently biasing the reconstructed tensor-to-scalar ratio $r$. Therefore, it becomes imperative to optimize the size of the spectral index pixels to ensure a bias smaller than the inherent sensitivity of the dataset.

We finally describe the component's frequency contribution with an operator by:
\begin{equation}
    \vec{A} = \begin{bmatrix}
        [\vec{A}^{\text{CMB}}_1, \vec{A}^{\text{Dust}}_1] & 0 & \cdots & 0 \\
        0 & [\vec{A}^{\text{CMB}}_2, \vec{A}^{\text{Dust}}_2] & & \vdots \\
        \vdots & & \ddots & 0 \\
        0 & \cdots & 0 & [\vec{A}^{\text{CMB}}_p, \vec{A}^{\text{Dust}}_p] \\
    \end{bmatrix},
    \label{eq:A_nu_varying}
\end{equation}
with $p$ the number of pixels in the HEALPix convention~\citep{Gorski_2005}. The number of reconstructed parameters increases computation time and reduces the signal-to-noise ratio within reconstructed sky pixels, so here again one needs to find the optimal number of parameters for a given dataset sensitivity.

\begin{figure*}
    \centering
    \resizebox{\hsize}{!}{\includegraphics[scale = 0.2]{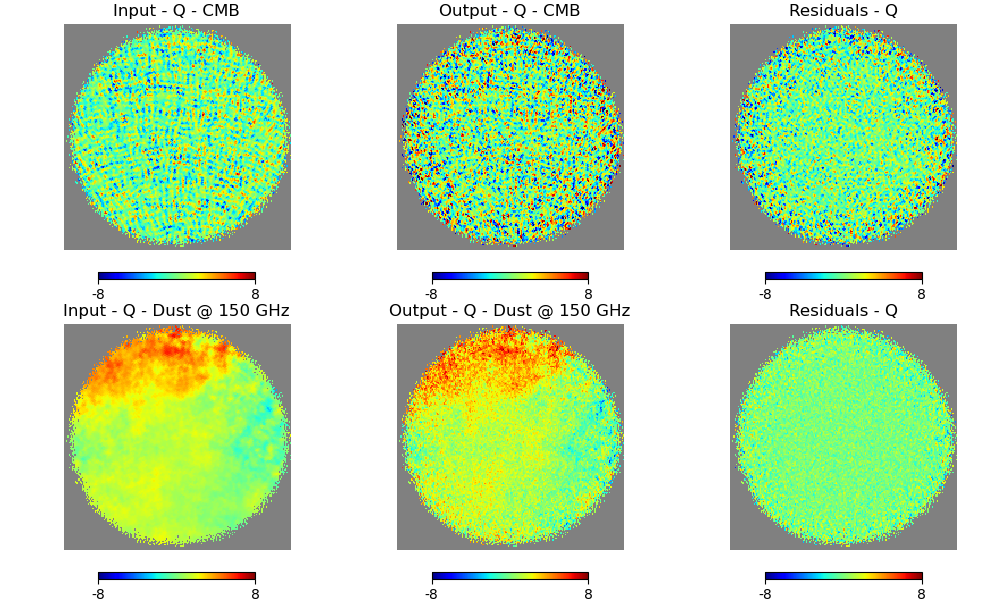}}
    \caption{Reconstruction of Q Stokes parameter, in $\si{\mu\K}_\text{CMB}$, for both component on the QUBIC patch (centered on $15^{\circ}$ radius sky patch at $\text{RA}=0^{\circ}$, $\text{DEC}=-57^{\circ}$) assuming varying spectral indices across the sky. Each row represents a component, and columns show the input, output, and residuals from left to right. We also include Planck HFI to constrain thermal dust emission.}
    \label{fig:components_varying}
\end{figure*}

\begin{figure}
    \centering
    \includegraphics[scale=0.4]{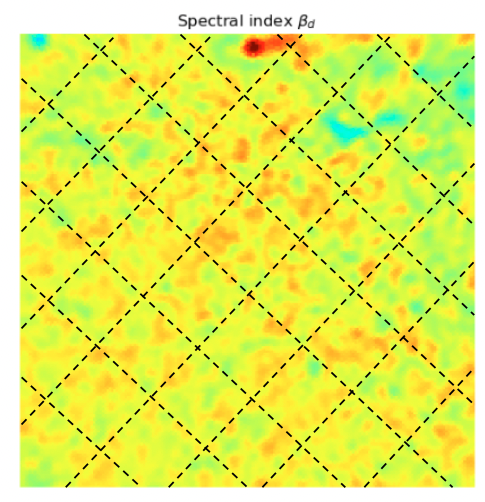}
    \caption{Illustration of the 2 pixelization scheme. Dashed diamonds show the resolution of the reconstructed spectral indices.}
    \label{fig: pix_scheme}
\end{figure}

In the same fashion as in the previous section, we achieve convergence of all unknowns through the alternate minimization scheme described in section~\ref{sec:review}. We emphasize that we use a double pixelization scheme for reconstruction with a fine \mbox{$N^{\text{sky}}_{\text{side}} = 256$} (corresponding to pixel with size about $13.74$ arcmin) for the foreground morphology and a coarse \mbox{$N_\text{side}^\beta = 8$} (corresponding to pixel with size about $7.3$ degrees) for spectral indices, meaning that the spectral index is assumed to be constant for all foreground map pixels lying within a coarse spectral index pixel. We show the double pixelization scheme in Fig.~\ref{fig: pix_scheme} where each color of the map shows a value of the spectral index for the input sky. The reconstruction is done using a coarser resolution shown by the black diamonds.

\begin{figure}
    \centering
    \includegraphics[scale=0.4]{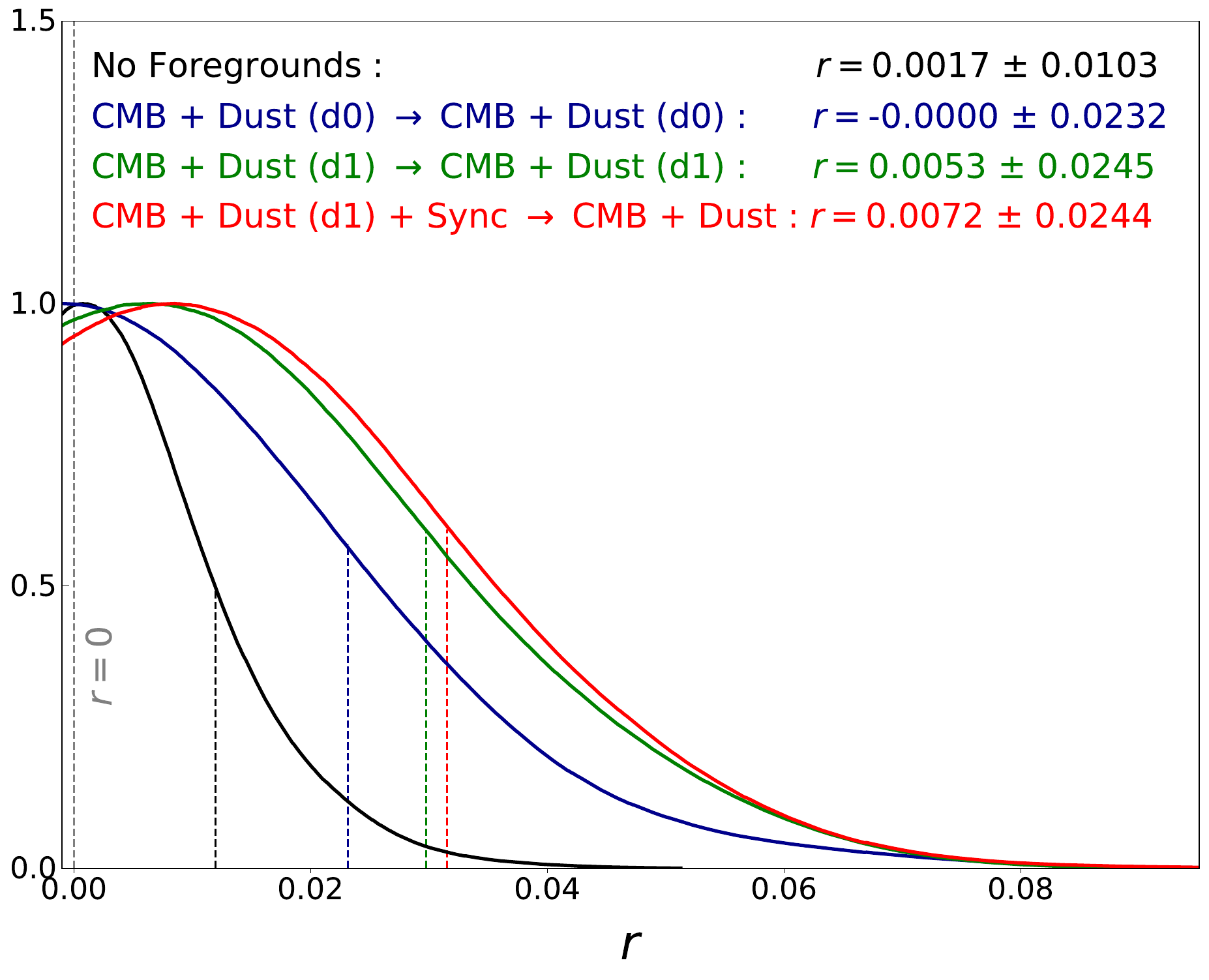}
    \caption{Posterior likelihood on tensor-so-scalar ratio $r$ assuming \textbf{d1} model.}
    \label{fig:cosmo}
\end{figure}

To finally conclude on these two methodologies, we conduct a cosmological analysis. Running 200 end-to-end simulations with identical skies but varying noise realizations, Fig.~\ref{fig:cosmo} shows the expected precision on tensor-to-scalar ratio $\sigma(r)$ for both scenarios. The blue curve reflects the findings detailed in Sec.~\ref{sec:constant_spectral_index}, while the green and red curves show the reconstructed $r$ in a scenario of varying spectral indices. Those 3 scenarios can be compared with the no foreground case shown by the solid black line. As anticipated we observe no sizable bias with \textbf{d0} while the spectral indices pixelization when analyzing the \textbf{d1} case induces a bias on $r$ at the level of $5\times 10^{-3}$ that is further increased to $7\times 10^{-3}$ when unaccounted synchrotron is included in the input sky. 
We observe a bias increase in the case of {\bf d1} model (spectral indices varying across the sky) due to the coarse pixelization we use for reconstruction (see Fig.~\ref{fig: pix_scheme}) that averages multiple spectral indices into a single large pixel, resulting in dust residuals in the clean CMB maps. When adding synchrotron in the input maps, but not attempting to reconstruct it (as it is very low in our frequency band), an additional bias is observed, but still below the width of the likelihood.
Regarding the forecasted sensitivity, no significant change in the reconstructed $\sigma(r)\sim 0.024$ is observed between these three cases.

\subsubsection{Monochromatic line reconstruction}
\label{sec: monochromatic line recon}

The forward model allows the reconstruction of diffuse components over a wide frequency range. However, there are also so-called "monochromatic" components, notably the carbon monoxide (CO) emission line within 220\, GHz band. Before targeting the cosmological field of QUBIC, the collaboration plan to scan the galactic plane to demonstrate QUBIC's unique spectral ability to distinguish frequencies.

In the past, missions like Planck have managed to create maps of these emissions~\citep{co_line_planck} by using the different bandpasses of each detector. However, in this article, we are exploring a new approach where we incorporate a model that focuses on a single line and calculates how it passes through our model at this specific frequency. This method allows us to better capture and study this emission in our TOD, fully benefiting from bolometric interferometry's specificity.

\begin{figure}
    \centering
    \includegraphics[scale=0.7]{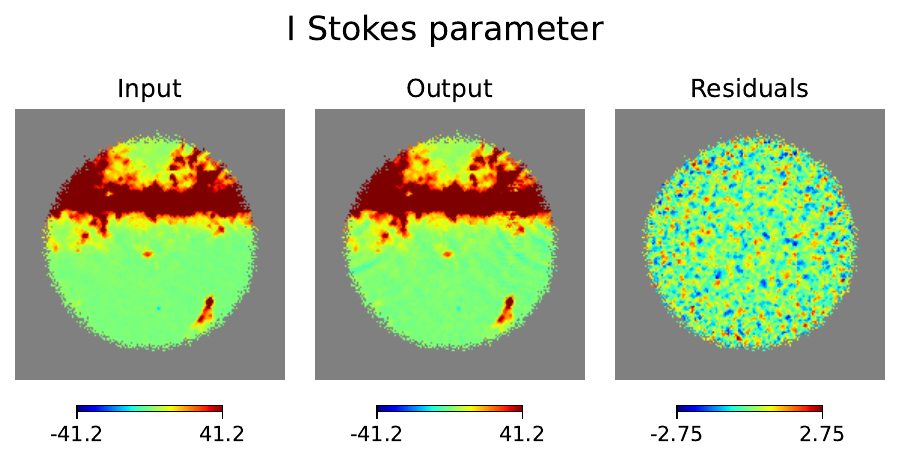}
    \caption{CO line reconstruction of I Stokes parameter, in $\si{\mu\K}_\text{CMB}$, using the monochromatic model on the galactic plane. From left to right, we show the input, output, and residuals, respectively.}
    \label{fig:co_line_I}
\end{figure}

\begin{figure}
    \centering
    \includegraphics[scale=0.7]{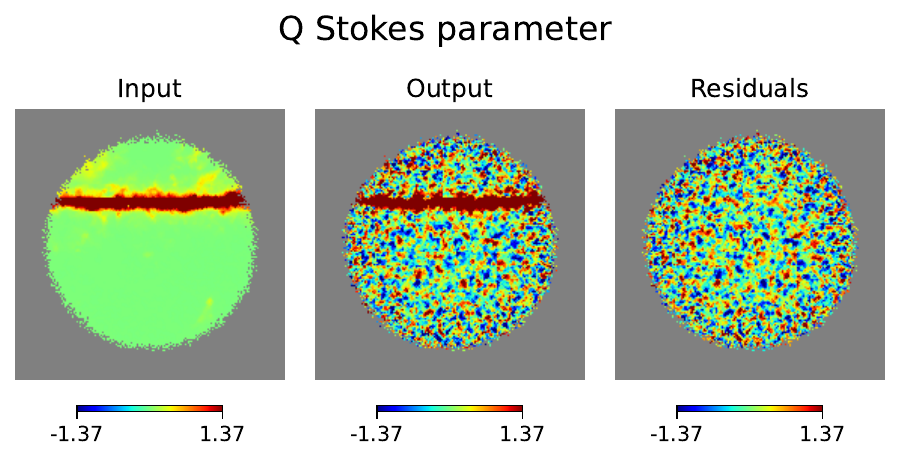}
    \caption{CO line reconstruction of Q Stokes parameter, in $\si{\mu\K}_\text{CMB}$, using the monochromatic model on the galactic plane. From left to right, we show the input, output, and residuals, respectively.}
    \label{fig:co_line_Q}
\end{figure}

In this study, our main goal is not to exploit a systematic effect for our benefit but to leverage the capabilities of spectral-imaging in distinguishing different frequencies. Specifically, we have observed that QUBIC has a synthesized beam that changes significantly with varying wavelengths. By having a good understanding of our PSF, we can calculate the beam at a specific frequency, allowing us to target a particular range of frequencies and extract a map of carbon monoxide emissions. Eq.~\ref{eq:hco} presents the updated reconstruction model (described in section~\ref{sec: sky_model}), which considers various diffuse emissions and incorporates a term for single-color emissions. It's important to note that we can reconstruct multiple line emissions by calculating the beam for each frequency of interest. Still, it is essential to know the specific frequency for each of these emissions. 

To reconstruct this particular component, we conducted a simulation involving three years of cumulative data acquisition along the Milky Way. We incorporate in the simulation primordial fluctuations contaminated both with thermal dust ({\bf d0} model) emission and CO line. To account for this monochromatic line, we add to the model the characteristics of the synthesized beam at a specific frequency, \mbox{$\nu_{\text{co}} = 230.538$~GHz}. This requires a accurate knowledge of the shape of the synthesized beam which can be achieved through direct mapping~\cite{2020.QUBIC.PAPER3} or through self-calibration~\cite{Bigot_Sazy_2013}. In the present study, we assume a perfect knowledge of the synthesized beam multiple peaks positions and amplitudes. 
We show the outcome of the reconstruction for the I and Q Stokes parameters, with Fig.~\ref{fig:co_line_I}, and Fig.~\ref{fig:co_line_Q} respectively. The reconstructed maps have been reconvolved by a Gaussian kernel to finally have an angular resolution of $\text{FWHM} = 0.5^{\circ}$. In the case of the polarization map, we assume a 1\% polarization fraction according to~\cite{Greaves_1999} stemming from the intensity map. This polarization fraction was chosen only for the sake of demonstrating the method. We note that incorporating this component into the analysis increases the overall computation time, as capturing this emission precisely is challenging, even with a detailed description of the synthesized beam at this frequency. Compared to the full bandwith of the intrument, very few photons will characterize this monochromatic emission. Therefore the noise level of this reconstruction will be very high.

\subsection{Blind component separation map-making}
\label{sec:blind}

\begin{figure*}
    \centering
    \includegraphics[scale=0.5]{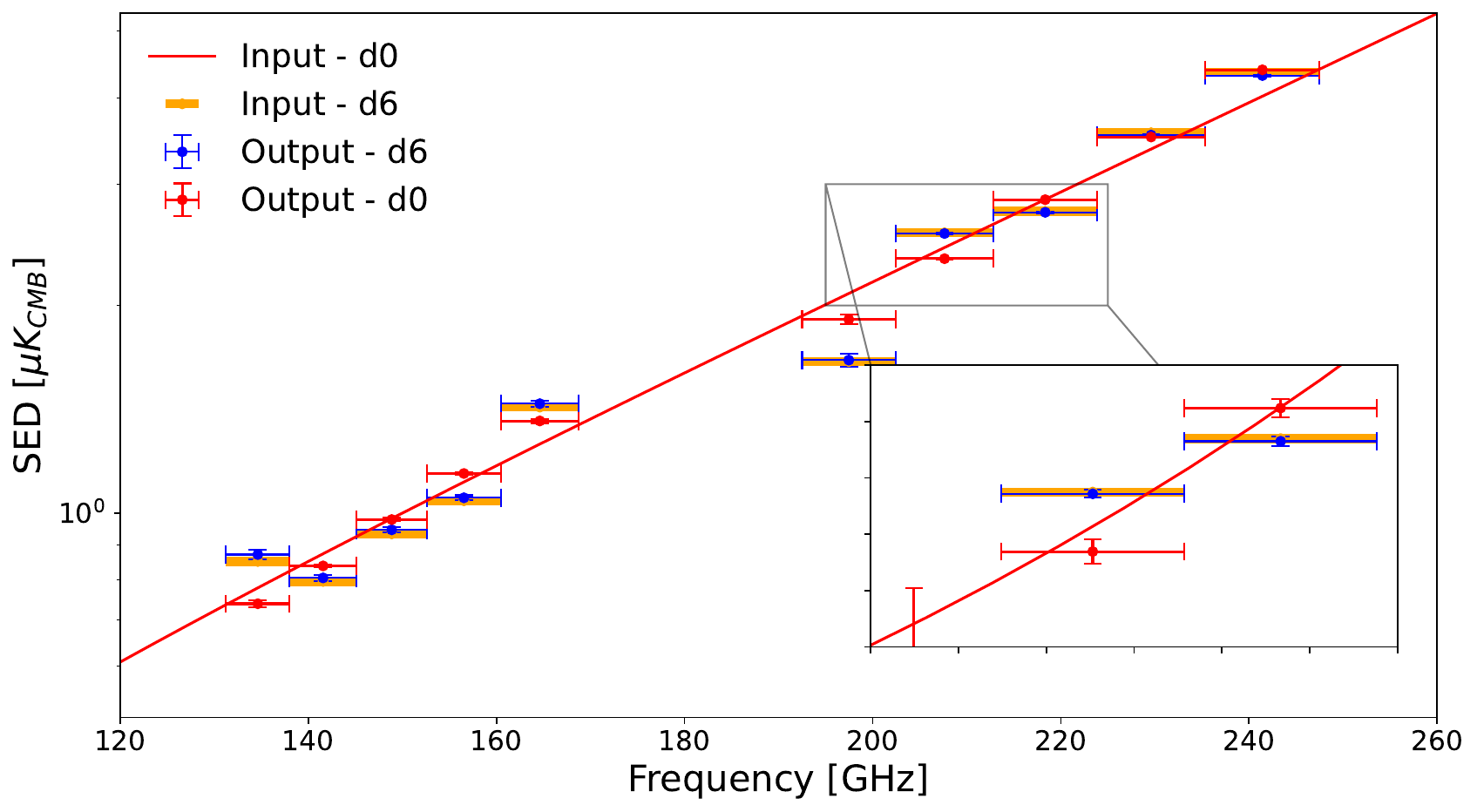}
    \caption{Reconstructed spectral energy distribution with ``blind'' method for simple and complex thermal dust emission. The box part shows a zoom on the first bin of the 220 GHz band. The horizontal bars show the bandpass of each sub-band considered in the reconstruction.}
    \label{fig:sed_blind}
\end{figure*}

In the pursuit of a more flexible and ‘‘data-driven'' approach, we extend our component separation algorithm to a ‘‘blind'' method, where we relax the assumption of an underlying global Spectral Energy Distribution (SED) model for the foregrounds. In this scenario, we no longer impose a predefined model for the mixing matrix elements, allowing the algorithm to automatically reconstruct an effective binned SED directly from the observed TOD.

The blind component separation approach proves valuable when foreground emission complexity is poorly characterized or varies significantly across the observed region \citep{regnier2023identifying}. As a last example, we can study the reconstruction of the components assuming primordial fluctuations contaminated with more complex thermal dust emission than \textbf{d0} and \textbf{d1}. We introduce the \textbf{d6} model, inspired from \cite{Vansyngel_2018} which includes frequency decorrelation of the dust. This is done by a randomization of the SED which mimics the effect of a frequency-varying polarization angle. This model is parametrized by $\ell_{\text{corr}}$ parameter which, as it decreases, increases dispersion around the MBB model meaning that a smaller value for $\ell_{\text{corr}}$ implies a stronger decorrelation in the dataset. In this method, we use a generalized component frequency contribution operator $A$ that doesn't rely on pre-established spectral indices. This adaptability is advantageous in situations where traditional methods may struggle to capture the diverse behaviors of foreground components.

The frequency discretization streamlines the reconstruction process by allowing a more direct exploration of the mixing matrix, which characterizes how each component contributes to each frequency channel, as shown in Fig.~\ref{fig:sed_blind}. At the moment, a unique spectrum can be estimated over the whole sky, implementing this method with the proper variation across the sky is the next step of this work.

\begin{figure}
    \centering
    \includegraphics[scale=0.4]{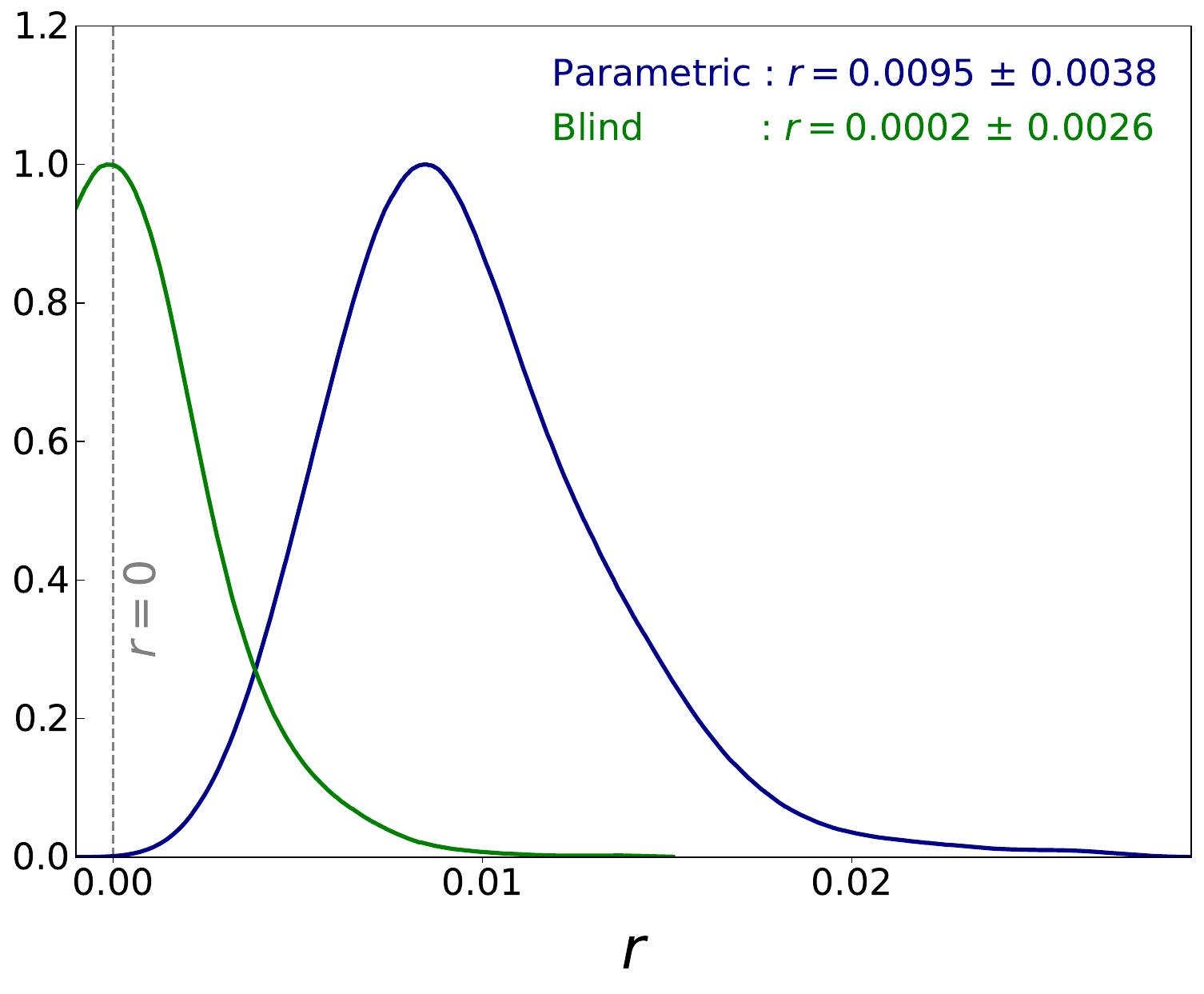}
    \caption{Posterior likelihood on tensor-to-scalar ratio $r$ for \textbf{d6} model. This result has been obtained with a hypothetical case with 10 times smaller noise.}
    \label{fig:cosmo_r_blind}
\end{figure}

Remarkably, even when thermal dust appears ostensibly simple, as in the case of the \textbf{d0} model represented by the solid red line, the method recovers the ``regular'' behavior of the SED as can be seen with the red points perfectly matching the underlying SED without being constrained to this shape.  The method's adaptability is further highlighted by its capacity to detect features in frequency space, exemplified by the detailed representation of the \textbf{d6} model in~\cite{Vansyngel_2018}.

Given QUBIC's anticipated sensitivity, the effect of dust decorrelation is not highly significant. To highlight Bolometric Interferometry's ability to constrain complex dust, we conducted end-to-end simulations with a hypothetical enhanced setup, reducing noise by a factor of 10 and assuming efficient delensing with lensing residuals at $A_\text{lens} = 0.5$. As the amount of synchrotron remains small in our bands~\citep{2020.QUBIC.PAPER1}, we decide to fix the spectral behavior of synchrotron emission, only reconstructing the amplitude along the LOS jointly with the CMB and thermal dust. This approach excels in accurately reconstructing decorrelated thermal dust emission, thereby mitigating biases in tensor-to-scalar ratio $r$ estimation, as illustrated in Fig~\ref{fig:cosmo_r_blind}.

The analysis is done using the same input data and almost the same reconstruction as before. The only difference is that instead of a parametric description of the $A$ mixing matrix elements (section~\ref{parametric}), we leave them as free parameters in the cost function minimization. We displayed the result in Fig.~\ref{fig:cosmo_r_blind} where the blue curve shows the reconstruction of posterior likelihood on $r$ in the case of a parametric method, revealing a biased tensor-to-scalar ratio due to decorrelated thermal dust emission. In contrast, the green curve reveals the results of the ``blind'' method, effectively removing the bias from decorrelated thermal dust. The blind approach removed the bias because of the more general approach that does not impose an SED model for astrophysical foregrounds. 

\section{Conclusion}

In this paper, we have shown how bolometric interferometry enables component separation at the map-making stage without the need to rebuild frequency maps. The method we developed, relying on spectral-imaging, is completely software-based and exploits the frequency dependence of the synthesized beam. 

This method is built through an inverse problem approach by modeling the instrumental setup and the astrophysical contributions producing the dataset. To achieve that, we also include Planck public maps, to regularize the edges in the same way as \cite{fmm}. Our method also natively accounts for the varying angular resolution of the instrument as a function of frequency. 

\renewcommand{\arraystretch}{1.6}
\begin{table}[h]
    \caption{Summary of reconstructed $r \pm \sigma(r)$.}              
    \label{table:1}      
    \centering                                      
    \begin{tabular}{| c | c | c | c |}          
        \hline                                  
        Input & Output & Parametric & Blind \\    
        \hline                                   
        \textbf{d0}   & \multirow{2}{1em}{\textbf{d0}} & 0.0000 $\pm$ 0.0232 & 0.0004 $\pm$ 0.0264\\      
        \textbf{d0s0} &                                & 0.0045 $\pm$ 0.0229 & 0.0021 $\pm$ 0.0262 \\
        
        \hline
        
        \textbf{d1}     & \multirow{2}{1em}{\textbf{d1}} & 0.0053 $\pm$ 0.0245 & \ding{53}     \\
        \textbf{d1s1}   &                                & 0.0072 $\pm$ 0.0244 & \ding{53} \\
        
        \hline
        
        \textbf{d6s0} & \textbf{d6s0} & 0.0095 $\pm$ 0.0038 & 0.0002 $\pm$ 0.0026 \\
        \hline                                             
    \end{tabular}
\end{table}

We produced end-to-end simulations to reconstruct astrophysical components directly from TODs, without the usual intermediate step of frequency maps. The latter enabled us to reconstruct the associated spectra and perform a cosmological analysis down to constraints on the tensor-to-scalar ratio $r$. This fit is made jointly with the spectral parameters of the dust on the separate maps to take advantage of the full noise covariance matrix. 

The instrumental setup we used is QUBIC Full Instrument, as described in \cite{2020.QUBIC.PAPER1} which considers two physical bands and 3 years of cumulative observations for each. For the noise contribution, we consider two components: detector noise fixed at $\text{NEP}_{\text{det}} = 4.7\times 10^{-17} \text{ } W/\sqrt{\text{Hz}}$ with 992 detectors per band, and photon noise. In this study, we explore the effect of astrophysical foregrounds to impact the measurement of the tensor-to-scalar ratio $r$, unlike \cite{2020.QUBIC.PAPER1} which assumed a pure CMB.

A summary of all the results is displayed in Table~\ref{table:1}. First, we have shown that this technique achieves a sensitivity on $r$ of the order of $\sigma(r) \sim 0.0229$ for a simplistic case where the foregrounds only consist of thermal dust following an MBB law with a constant spectral index (model \textbf{d0}) across the sky. When including a similarly constant spectral index synchrotron component (model \textbf{d0s1}), a non-negligible bias remains around $r \sim 0.0045$, because of the impossibility of reconstructing the latter due to the lack of low-frequency bands with the instrumental configuration considered here. 

In a second step, we extended the method to a case where the spectral indices of the dust vary according to the LOS (model \textbf{d1}). This method is more computationally demanding, forcing us to reduce the pixelization of spectral index reconstruction to $N_\text{side}^\beta = 8$ (unlike the components amplitude maps, which remain at $N_\text{side}^\text{sky} = 256$). These simulations show a similar error on $r$ with a slight increase of the bias due to variation in these spectral indices to $r \sim 0.0072$.

Finally, we have further extended the method towards blind reconstruction. Using the frequency description of the synthesized beam, we can minimize a cost function that is not based on a parametric description, but directly on the elements of the mixing matrix for the foregrounds. This is made possible, here again, by the frequency variation of QUBIC's synthetized beam, which encodes spatially the frequency evolution of the incoming radiation. Simulations are based on \cite{regnier2023identifying}, reproducing the effect of thermal dust frequency decorrelation (model \textbf{d6s0}). As expected, parametric reconstruction leads to a significant bias around \mbox{$r_{\text{bias}} \sim 0.01$} on r due to significant dust residuals in the CMB clean map. These completely disappear when the blind method is used. This new approach not only provides a more efficient component separation but also makes it possible to characterize astrophysical foregrounds and their complexities at a small frequency scale. 

In this study, we have assumed a stable atmosphere, which simplifies our analysis. Ongoing efforts aim to refine the instrumental model and remove atmospheric contributions from the data using spectral-imaging techniques. Additionally, we plan to incorporate a more comprehensive set of instrumental systematics. Optimization work is also in progress to apply the blind method to account for potential variations across the sky.

\acknowledgments
QUBIC is funded by the following agencies. France: ANR (Agence Nationale de la Recherche) contract ANR-22-CE31-0016, DIM-ACAV (Domaine d’Intérêt Majeur-Astronomie et Conditions d’Apparition de la Vie), CNRS/IN2P3 (Centre national de la recherche scientifique/Institut national de physique nucléaire et de physique des particules), CNRS/INSU (Centre national de la recherche scientifique/Institut national et al de sciences de l’univers). Italy: CNR/PNRA (Consiglio Nazionale delle Ricerche/Programma Nazionale Ricerche in Antartide) until 2016, INFN (Istituto Nazionale di Fisica Nucleare) since 2017.  Argentina: MINCyT (Ministerio de Ciencia, Tecnología e Innovación), CNEA (Comisión Nacional de Energía Atómica), CONICET (Consejo Nacional de Investigaciones Científicas y Técnicas).



\bibliographystyle{JHEP}
\bibliography{biblio.bib}

\appendix

\begin{appendix}

\section{Dectectors relative gain}
\label{appendixA}

In our article, we produce an alternate estimation of components amplitude, spectral indices, and systematics. In this specific study, we include systematics, the gain of each detector, as an example. We derive the semi-analytical solution using the expression of measured data by:
\begin{equation}
    \vec{d}_a = g_a \vec{H}_a \vec{A} \vec{c} + \vec{n},
    \label{eq:data_measured}
\end{equation}
where $g_a$ is the intercalibration factor for the detector $a$. We will omit the indices $a$ for the rest because we focus only on one detector. We are using an alternate approach, meaning that for the $i$-th iteration, an estimation of the component vector $\vec{c}$ and mixing matrix $\vec{A}$ is available. The model assumes a constant gain as a function of time samples, but more refined models can be implemented. We can, therefore, build a cost function to minimize for the $i$-th iteration:
\begin{equation}
\begin{split}
    \chi^2_i &= \left( \vec{d} - \vec{\Tilde{d}}_i \right)^T \vec{N}^{-1} \left( \vec{d} - \vec{\Tilde{d}}_i \right) \\
           &= \left( \vec{d} - \Tilde{g}_i \vec{H} \vec{\Tilde{A}}_i \vec{\Tilde{c}}_i \right)^T \vec{N}^{-1} \left( \vec{d} - \Tilde{g}_i \vec{H} \vec{\Tilde{A}}_i \vec{\Tilde{c}}_i \right) \\
           &= \vec{d}^T \vec{N}^{-1} \vec{d} - 2 \vec{\Tilde{c}}^T \vec{\Tilde{A}}^T \vec{H}^T \Tilde{g}_i^T \vec{N}^{-1} \vec{d} +
             \vec{\Tilde{c}}_i^T \vec{\Tilde{A}}_i^T \vec{H}^T \Tilde{g}_i^T \vec{N}^{-1} \Tilde{g}_i \vec{H} \vec{\Tilde{A}}_i \vec{\Tilde{c}}_i.
\end{split}
\end{equation}

We find the minimium with respect to $\Tilde{g}_i^T$ with:
\begin{equation}
\begin{split}
    \nabla_{\Tilde{g}_i\,} \chi^2_i & = - 2 \vec{\Tilde{c}}_i^T \vec{\Tilde{A}}_i^T \vec{H}^T \vec{N}^{-1} \vec{d} + 2 \vec{\Tilde{c}}_i^T \vec{\Tilde{A}}_i^T \vec{H}^T \vec{N}^{-1} \Tilde{g}_i \vec{H} \vec{\Tilde{A}}_i \vec{\Tilde{c}}_i \\
    & = 0.
\end{split}
\end{equation}

\begin{figure}
    \centering
    \includegraphics[scale=0.60]{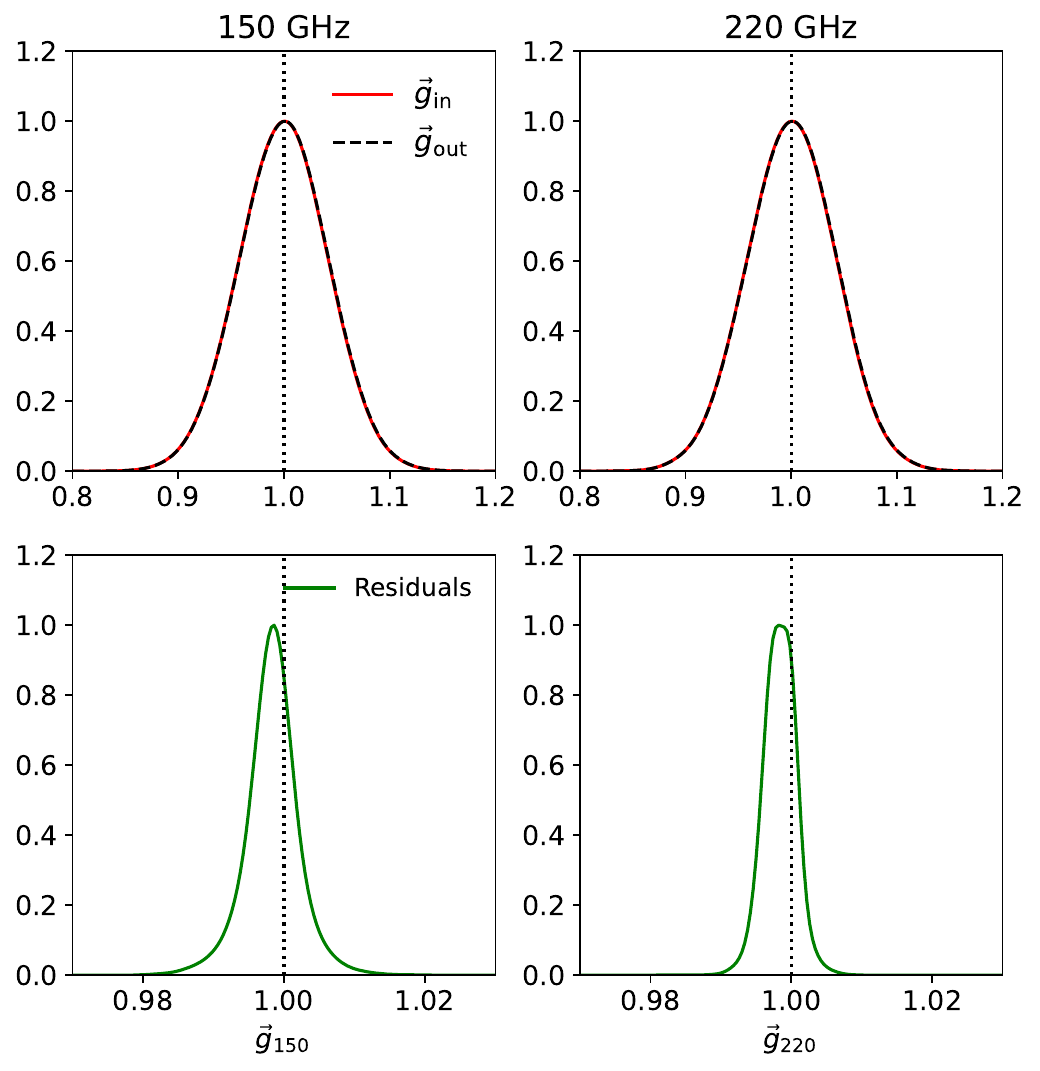}
    \caption{\textbf{(Top-panel):} Distribution of input (red solid lines) and reconstructed (dashed black lines) for both 150 and 220 \,GHz. \textbf{(Bottom-panel):} Distribution of residuals for each frequency.}
    \label{fig: hist_gain}
\end{figure}

Knowing that $\vec{\Tilde{d}}_i = \vec{H} \vec{\Tilde{A}}_i \vec{\Tilde{c}}_i$ is our simulated data at iteration $i$ assuming perfect intercalibration, we can finally derive:
\begin{equation}
    \Tilde{g}_i = \frac{\vec{\Tilde{d}}_i^T \vec{N}^{-1} \vec{d}}{\vec{\Tilde{d}}_i^T \vec{N}^{-1} \vec{\Tilde{d}}_i},
\end{equation}
where $\Tilde{g}_i$ is an intercalibration factor for the detector $a$. An example of the accuracy of the reconstruction is displayed in Fig.~\ref{fig: hist_gain}, showing the histograms of the residuals for both focal planes. Most of the intercalibration factors can be recovered with a precision smaller than $10^{-2}$ for both focal planes. The method we present in this article can be used to estimate systematics as the gain detector during the map-making process. However, bolometric interferometry allows to use of several ways of estimating instrumental systematics, as the self-calibration phase \cite{Bigot_Sazy_2013}. The method introduced in this article is complementary to the others; systematics can be estimated in another way and then fixed during the components map-making. 

\section{Resolution of the final maps without convolutions during reconstruction} \label{sec: appendixB}

As explained in section~\ref{spectral_imaging}, we do the reconstruction with a simulated TOD written as:
\begin{equation}
    \vec{\Tilde{d}} = \left( \sum_{k = 1}^{N_{\text{sub}}}
    \Delta\nu_k \mathcal{H}_{\nu_k} \mathcal{C}_{K_k} \vec{\Tilde{A}}_{\nu_k} \right) \cdot \vec{\Tilde{c}}.
\end{equation}
This is computationally very expensive because of the convolutions operators $\mathcal{C}_{K_k}$. So we want to explore the possibility of not doing the convolutions during the reconstruction. The real TOD is still generated through convoluted maps, so the PCG will try to fit components maps $\vec{\Tilde{c}} = \left(\vec{\Tilde{s}}_\text{CMB}^{\sigma_\text{CMB}},\, \vec{\Tilde{s}}_\text{dust}^{\sigma_\text{dust}},\, \vec{\Tilde{s}}_\text{sync}^{\sigma_\text{sync}}\right)$ at some resolutions $\sigma_\text{CMB}$, $\sigma_\text{dust}$, $\sigma_\text{sync}$, common for the $N_\text{sub}$ frequencies of the band. The simulated TOD is then:
\begin{equation}
    \vec{\Tilde{d}} = \left( \sum_{k=1}^{N_\text{sub}} \Delta\nu_k \mathcal{H}_{\nu_k} \vec{\Tilde{A}}_{\nu_k} \right) \cdot \left(\vec{\Tilde{s}}_\text{CMB}^{\sigma_\text{CMB}},\, \vec{\Tilde{s}}_\text{dust}^{\sigma_\text{dust}},\, \vec{\Tilde{s}}_\text{sync}^{\sigma_\text{sync}}\right).
\end{equation}
We want to determine $\sigma_\text{CMB}$, $\sigma_\text{dust}$, $\sigma_\text{sync}$. The PCG fits the simulated TOD to the real TOD, so we get the equality:
\begin{multline}
    \left( \sum_{k=1}^{N_\text{sub}} \Delta\nu_k \mathcal{H}_{\nu_k} \vec{\Tilde{A}}_{\nu_k} \right) \cdot \left(\vec{\Tilde{s}}_\text{CMB}^{\sigma_\text{CMB}},\, \vec{\Tilde{s}}_\text{dust}^{\sigma_\text{dust}},\, \vec{\Tilde{s}}_\text{sync}^{\sigma_\text{sync}}\right) 
    \approx \\
    \left( \sum_{k=1}^{N_\text{sub}} \Delta\nu_k \mathcal{H}_{\nu_k} \mathcal{C}_{\sigma_k} \vec{A}_{\nu_k} \right) \cdot \left(\vec{s}_\text{CMB}^\infty,\, \vec{s}_\text{dust}^\infty,\, \vec{s}_\text{sync}^\infty\right),
\end{multline}
\begin{multline}
    \left(\vec{\Tilde{s}}_\text{CMB}^{\sigma_\text{CMB}},\, \vec{\Tilde{s}}_\text{dust}^{\sigma_\text{dust}},\, \vec{\Tilde{s}}_\text{sync}^{\sigma_\text{sync}}\right) \approx 
    \left( \sum_{k=1}^{N_\text{sub}} \Delta\nu_k \mathcal{H}_{\nu_k} \vec{\Tilde{A}}_{\nu_k} \right)^{-1} \cdot \\
    \left( \sum_{k=1}^{N_\text{sub}} \Delta\nu_k \mathcal{H}_{\nu_k} \vec{A}_{\nu_k} \cdot \left(\mathcal{C}_{\sigma_k} \vec{s}_\text{CMB}^\infty,\, \mathcal{C}_{\sigma_k} \vec{s}_\text{dust}^\infty,\, \mathcal{C}_{\sigma_k} \vec{s}_\text{sync}^\infty\right) \right).
\end{multline}
This means that the reconstructed maps $\left(\vec{\Tilde{s}}_\text{CMB}^{\sigma_\text{CMB}},\, \vec{\Tilde{s}}_\text{dust}^{\sigma_\text{dust}},\, \vec{\Tilde{s}}_\text{sync}^{\sigma_\text{sync}}\right)$ are the average of the maps $\left(\mathcal{C}_{\sigma_k} \vec{s}_\text{CMB}^\infty,\, \mathcal{C}_{\sigma_k} \vec{s}_\text{dust}^\infty,\, \mathcal{C}_{\sigma_k} \vec{s}_\text{sync}^\infty\right)$ weighted by the operators $\Delta\nu_k \mathcal{H}_{\nu_k} \vec{A}_{\nu_k}$. It's not very practical to weight an average with operators, so we introduce the scalars $h_{\nu_k}$:
\begin{equation}
    h_{\nu_k} = \mathcal{H}_{\nu_k} \cdot \mathcal{I},
\end{equation}
which represent the scale of the operators $\mathcal{H}_{\nu_k}$ and $\mathcal{I}$ is a uniform sky set to one everywhere. Then we can write the reconstructed maps as:
\begin{equation}
\begin{split}
    & \vec{\Tilde{s}}_\text{CMB}^{\sigma_\text{CMB}} \approx \dfrac{\sum_{k=1}^{N_\text{sub}} \Delta\nu_k h_{\nu_k} \mathcal{C}_{\sigma_k} \vec{s}_\text{CMB}^\infty}{\sum_{k=1}^{N_\text{sub}} \Delta\nu_k h_{\nu_k}} = \left\langle \mathcal{C}_{\sigma_k} \right\rangle_k \vec{s}_\text{CMB}^\infty \\
    & \vec{\Tilde{s}}_\text{dust}^{\sigma_\text{dust}} \approx \dfrac{\sum_{k=1}^{N_\text{sub}} \Delta\nu_k h_{\nu_k} f_d^{\,\beta_d}(\nu_k) \mathcal{C}_{\sigma_k} \vec{s}_\text{dust}^\infty}{\sum_{k=1}^{N_\text{sub}} \Delta\nu_k h_{\nu_k} f_d^{\,\beta_d}(\nu_k)} = \left[ \mathcal{C}_{\sigma_k} \right]_k \vec{s}_\text{dust}^\infty \\
    & \vec{\Tilde{s}}_\text{sync}^{\sigma_\text{sync}} \approx \dfrac{\sum_{k=1}^{N_\text{sub}} \Delta\nu_k h_{\nu_k} f_s^{\,\beta_s}(\nu_k) \mathcal{C}_{\sigma_k} \vec{s}_\text{sync}^\infty}{\sum_{k=1}^{N_\text{sub}} \Delta\nu_k h_{\nu_k} f_s^{\,\beta_s}(\nu_k)} = \left\{ \mathcal{C}_{\sigma_k} \right\}_k \vec{s}_\text{sync}^\infty,
\end{split}
\end{equation}
where we introduced the notations $\langle . \rangle_k$, $[ . ]_k$, $\{ . \}_k$ which are the average over $k$ weighted by $\Delta\nu_k h_{\nu_k}$, $\Delta\nu_k h_{\nu_k} f_d^{\,\beta_d}(\nu_k)$ and $\Delta\nu_k h_{\nu_k} f_s^{\,\beta_s}(\nu_k)$ respectively.

We can approximate $\left\langle \mathcal{C}_{\sigma_k} \right\rangle_k$, $\left[ \mathcal{C}_{\sigma_k} \right]_k$ and $\left\{ \mathcal{C}_{\sigma_k} \right\}_k$ by convolutions with Gaussian functions of width:
\begin{equation}
\begin{split}
    & \sigma_\text{CMB} = \left\langle \sigma_k \right\rangle_k \\
    & \sigma_\text{dust} = \left[ \sigma_k \right]_k \\
    & \sigma_\text{sync} = \left\{ \sigma_k \right\}_k.
\end{split}
\end{equation}
These are the resolutions of the maps after the reconstruction without the convolutions during the process.

For the parametric method, the reconstruction without convolutions appears to behave well. But for the blind method, the reconstruction without convolutions seems to have difficulty to reconstruct the mixing matrix $\vec{\Tilde{A}}_\nu$. Further investigation is needed.
\end{appendix}

\end{document}